\documentclass[usenatbib]{mn2e}
\usepackage{graphicx}
\usepackage{ifthen}
\usepackage{url}
\usepackage{fixltx2e}

\def\ltsima{$\; \buildrel < \over \sim \;$}
\def\lta{\lower.5ex\hbox{\ltsima}}
\def\gtsima{$\; \buildrel > \over \sim \;$}
\def\simgt{\lower.5ex\hbox{\gtsima}}
\def\kms{{\rm\,km\,s^{-1}}}

\def\kpc{{\rm\,kpc}}

\def\msun{{\rm\,M_\odot}}
\def\lsun{{\rm\,L_\odot}}

\def\Aa{\; \buildrel \circ \over {\rm A}}
\def\AA{$\; \buildrel \circ \over {\mathrm A}$}

\def\deg{^\circ}

\def\s{\ifmmode \widetilde \else \~\fi}
\def\={\overline}

\def\spose#1{\hbox to 0pt{#1\hss}}

\def\lta{\mathrel{\spose{\lower 3pt\hbox{$\mathchar"218$}}
     \raise 2.0pt\hbox{$\mathchar"13C$}}}
\def\gta{\mathrel{\spose{\lower 3pt\hbox{$\mathchar"218$}}
     \raise 2.0pt\hbox{$\mathchar"13E$}}}
\def\Dt{\spose{\raise 1.5ex\hbox{\hskip3pt$\mathchar"201$}}}    
\def\dt{\spose{\raise 1.0ex\hbox{\hskip2pt$\mathchar"201$}}}    

\def\dotsfill{\leaders\hbox to 1em{\hss.\hss}\hfill}

\def\feh{{\rm[Fe/H]}}
\def\afe{{\rm[\alpha/Fe]}}

\def\ec4{EC4}

\loadboldmathitalic 

\title[The thick disc in M31]{The kinematic identification of a thick stellar disc in M31$^{1,2}$}
\author [M.\ L.\ M.\ Collins et al.] {M. L. M. Collins$^3$,
  S. C. Chapman$^3$, R. A. Ibata$^4$, M. J. Irwin$^3$, R. M.
  Rich$^5$, \newauthor A. M. N. Ferguson$^6$, G. F. Lewis$^7$,
  N. Tanvir$^8$, A. Koch$^8$\\ $^3$Institute of Astronomy,Madingley
  Rise, Cambridge, CB3 0HA ,UK\\ $^4$ Observatoire de Strasbourg,11,
  rue de l'Universit\'e, F-67000, Strasbourg, France\\ $^5$ Department
  of Physics and Astronomy, University of California, Los Angeles, CA
  90095-1547 \\ $^6$ Institute for Astronomy, University of Edinburgh,
  Royal Observatory, Blackford Hill, Edinburgh, UK EH9 3HJ\\ $^7$
  Sydney Institute for Astronomy, School of Physics, A29, University
  of Sydney, NSW 2006, Australia\\ $^8$ Department of Physics \&
  Astronomy, University of Leicester, University Road, Leicester LE1
  7RH, UK\\} \date{\today}
\begin{document}
\maketitle 

\begin{abstract}
We present the first characterization of a thick disc component in the
Andromeda galaxy (M31) using kinematic data from the DEIMOS
multi-object spectrograph instrument on Keck II. Using 21 fields in
the South West of the galaxy, we measure the lag of this component
with respect to the thin disc, as well as the dispersion, metallicity
and scale length of the component. We find an average lag between the
two components of $\langle\Delta v\rangle=46.0\pm3.9\kms$. The
velocity dispersion of the thick disc is
$\sigma_{thick}=50.8\pm1.9\kms$, greater than the value of dispersion
we determine for the thin disc, $\sigma_{thin}=35.7\pm1.0\kms$. The
thick disc is more metal poor than the thin disc, with
[Fe/H]$_{spec}=-1.0\pm0.1$ compared to [Fe/H]$_{spec}=-0.7\pm0.05$ for
the thin disc. We measure a radial scale length of the thin and thick
discs of $h_r=7.3\pm1.0$~kpc and $h_r=8.0\pm1.2$~kpc. From this, we
infer scale heights for both discs of 1.1$\pm0.2$~kpc and 2.8$\pm0.6$
kpc, both of which are $\sim$2--3 times larger than those observed in
the Milky Way. We estimate a mass range for the thick disc component of
2.4$\times10^{10}\msun<M_{*,thick}<4.1\times10^{10}\msun$. This value
provides a useful constraint on possible formation mechanisms, as any
proposed method for forming a thick disc must be able to heat (or
deposit) at least this amount of material.
\end{abstract}

\section{Introduction}
\footnotetext[1]{The data presented herein were obtained at the
  W.M. Keck Observatory, which is operated as a scientific partnership
  among the California Institute of Technology, the University of
  California and the National Aeronautics and Space
  Administration. The Observatory was made possible by the generous
  financial support of the W.M. Keck Foundation.}
  \footnotetext[2]{
  Based on observations obtained with MegaCam, a joint
  project of CFHT and CEA/DAPNIA, at the Canada-France-Hawaii
  Telescope (CFHT) which is operated by the National Research Council
  (NRC) of Canada, the Institute National des Sciences de l'Univers of
  the Centre National de la Recherche Scientifique of France, and the
  University of Hawaii.}

Roughly 70\% of bright galaxies observed at redshift $z=0$ possess
stellar discs (e.g. \citealt{hammer05,park07,choi07} and
\citet{dserrano10}), including our own Galaxy and its two largest
neighbours, M31 and M33. From this, we can infer that spiral
morphologies are the dominant configuration for galaxies viewed at the
present epoch. Under the formalism of hierarchical structure
formation, galaxies are believed to evolve into their present forms
via the accretion of, and mergers with, smaller systems. The effect of
this process on the seemingly dynamically stable stellar discs we see
in Milky Way (MW) type galaxies is still largely uncertain. The
ability of these fragile objects to survive a ``major merger'' event
(i.e. a merger with a system $>1/3$ of the host's mass) is something
that is still debated. Such major mergers are thought to be
cosmologically common, with $\sim70\%$ of all galaxies with a halo
mass of $M\sim10^{12}\msun$ having experienced at least one major
interaction within the past 8 Gyrs \citep{stewart08,purcell09}. Thus
it has been argued that galaxies that possess thin stellar discs at
$z=0$ could not have experienced a major merger within the last 10 Gyr
without the disc being destabilized
\citep{toth92,walker96,stewart08,purcell09}. This poses a significant
challenge to our understanding of the formation of disc galaxies like
the MW and M31. Recently, several authors have argued that these thin
discs could survive such an event if the merging system is
sufficiently gas rich
\citep{robertson06,brook07,hopkins09,stewart09,brooks09}, although the
disc would still undergo heating, resulting in a thicker disc than
that observed presently in the MW.  In addition to the effect of major
mergers on the structure of discs, galaxies viewed at the present
epoch have undergone (and are still undergoing) many smaller ``minor''
mergers which are not sufficiently massive to destroy thin stellar
discs, but are thought to kinematically heat them, causing them to
flare outwards and create a second, thick disc component
\citep{quinn93,robin96,walker96,velazquez99,chen01,sales09,villalobos09,purcell10}.
Other physical processes are also thought to heat up and thicken the
thin disc, including the accretion of a satellite on a radial orbit
about its host \citep{abadi03,read08}, internal heating within the
disc from massive star clusters, interactions with spiral arms,
etc. (\citealt{villumsen85,carlberg87,sellwood02,hanninen02,benson04,hayashi06,kazantzidis07,roskar08,schonrich09a,loebman10}). Thick
discs may also have formed thick, with significant star formation
occurring above the mid-plane of the galaxy or with large initial
velocity dispersions \citep{brook04,kroupa02}. In recent work by
\citet{roskar10}, they suggest that in-situ formation
could also occur if the stellar disc is misaligned with the hot, gaseous
halo. This misalignment results in a significant warping of the
outer disc, and subsequent star formation within this warp results in a low
metallicity thick disc. Finally, it is also possible that a number of
these mechanism will act in conjunction. In particular, it has been
suggested by a number of authors that secular growth from internal
heating may be significantly enhanced by minor merger events via swing
amplification (e.g. \citealt{sellwood98,dubinski08}), as these
processes often occur simultaneously. As such, it makes little sense
to treat these two scenarios as separate processes. 

With so many potential mechanisms capable of producing thickened
stellar discs, just how common are thick discs in spiral galaxies at
the present epoch?  \citet{dalcanton02} claim that thick disc
formation is a universal feature of disc formation, and as such should
be observed in all spiral galaxies. Whether such discs are formed
predominantly via one mechanism, or a mixture of them is still
uncertain, and disentangling the various formation scenarios from one
another in present data sets has proven difficult.

In the MW, the existence of a thick disc has long been known, and was
first identified by \citet{gilmore83}. Subsequent spectroscopic
studies of this component have shown it to be kinematically distinct
from the thin stellar disc, with the thick disc lagging behind the
thin disc by $\sim50\kms$ \citep{carollo10} and having a larger
velocity dispersion than the thin disc. This thick component also
seems to be composed of older, more metal deficient stars
(e.g. \citealt{chiba00,wyse06}). However the observed properties of
the thick disc, such as scale height, length and velocity dispersion,
tend to vary depending on the survey sample and tracer population used
\citep{juric08,ivezic08,carollo10,dejong10}. As such, the origin of
the MW thick disc is still a subject of great debate in the
literature. Thick discs have also been observed in a number of edge on
spiral galaxies
(e.g. \citealt{burstein79,tsikoudi79,vanderkruit84,shaw89,vandokkum94,dalcanton02,elmegreen06,yoachim06,yoachim08a,yoachim08b}),
and spectroscopic observations of these objects also show the thick
discs to be composed of older stars than their corresponding thin
discs. However, as these galaxies are all located at distances greater
than $\sim$10 Mpc from the MW, one cannot obtain spectra for
individual stars, and must instead rely on the integrated spectral
properties of RGB stars. Obtaining spectra with a high enough
signal-to-noise (S:N) to discern velocity dispersion profiles and
reliable metallicities is also challenging, making it impossible to
distinguish between the various formation mechanisms for these
structures.

If thick stellar discs are universal amongst spiral galaxies, and are
formed by mergers with, or accretions of, satellites, one might expect
to see such a structure in M31. This neighbouring galaxy is considered
to be a ``typical'' spiral galaxy when compared with other local
external disc galaxies \citep{hammer07}. It is thought to have had an
active merger history, and a recent panoramic photometric survey by
the Pan-Andromeda Archaeological Survey collaboration (PAndAS,
\citealt{mcconnachie09}) has shown the halo of this galaxy to be
littered with tidal streams from interactions with in-falling
satellites. These include the Giant Southern Stream (GSS,
\citealt{ibata01b,gilbert09}) and streams A, B, C, D and E
\citep{ibata07,chapman08,mcconnachie09}. The outer disc of M31 is very
perturbed \citep{ferguson02,richardson08}, suggestive of some tidal
interaction. M31 is also host to 25 known dSph and 4 dE satellites, at
least 2 of which (NGC 205 and M32) show evidence for significant tidal
interaction \citep{choi02,mcconnachie04a,geha06,howley08}. Therefore
the possibility of numerous interactions between the disc of M31 and
its satellite population seems highly likely.  \citet{mcconnachie09}
also present evidence for a recent interaction between M31 and its
neighbouring spiral galaxy, M33, which could have significantly
distorted and heated the M31 disc, giving rise to a thick disc
component or substantial substructure in the outer disc.  Other groups
have postulated links between the formation of bulges and thick discs
in spiral galaxies \citep{melendez08,hopkins08,bournaud09,bensby10},
and as M31 is known to have a reasonably massive bulge
\citep{saglia10}, it is an interesting candidate for hosting a thick
disc. Despite its high inclination to us along the line of sight
(77$\deg$, \citealt{walterbos88}), M31 is not seen sufficiently
edge-on to allow us to look for such a population using photometry.
Therefore to look for evidence of a thick disc in M31, we must search
for it via its kinematic signature, using spectroscopy. Given its
proximity to us (785 kpc, \citealt{mcconnachie05a}), M31 is an ideal
target for spectroscopic observations as we are able to resolve and
obtain reliable velocities for individual Red Giant Branch (RGB)
stars, and it has an advantage over our own galaxy as we are afforded
a panoramic view, whereas in the MW we are hampered along various
lines of sights by confusion from the disc and the bulge.

Since 2002 our group has been conducting a systematic spectroscopic
survey of M31, including the disc, halo and regions of substructure
using the DEIMOS instrument mounted on the Naysmyth focus of Keck II
(I05,
\citealt{chapman05,chapman06,chapman07,chapman08,collins09,collins10}). The
data from this survey gives us an ideal opportunity to identify a
thick disc if present. In fact, in their study of M31's extended disc
using this same data set, I05 identified a population lagging behind
the thin disc which they excluded from their study that they termed a
`thick disc-like' population and \citet{chapman06} briefly examined
this component as a function of radius but were unable to comment on
its global properties. In this work, we discuss the results from an in
depth study of this population, analysing its kinematics and
chemistry, comparing them to M31's thin and extended discs, the thin
and thick discs in the MW, and those seen in other galaxies, and
comment on possible formation scenarios for this component. The paper
is set out as follows; in \S2 we discuss the known structure of M31,
\S3 focuses on our spectroscopic survey of M31 and discusses field
selection and analysis techniques. We present our results in \S4 and
discuss their implications in \S5. We conclude our findings in \S6.
\begin{figure}
\begin{center}
\includegraphics[angle=0,width=0.99\hsize]{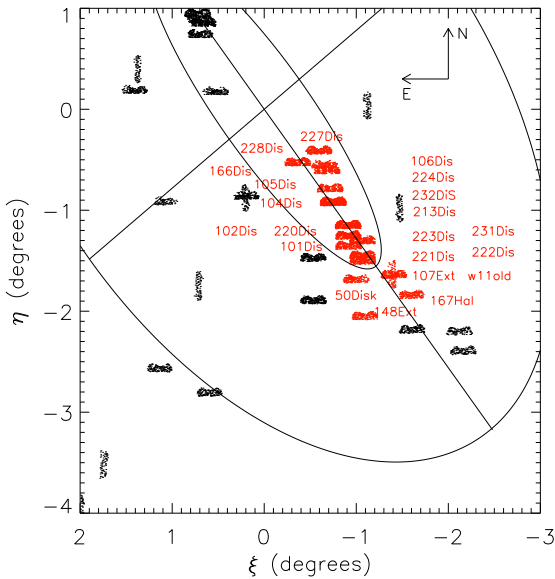}
\caption{Map showing the location of fields within our M31
  survey. Fields selected for study in this work are labelled and
  highlighted in red. The outer ellipse shows a segment of a 55 kpc
  radius ellipse flattened to c/a= 0.6, and the major and minor axis
  are indicated with straight lines out to this ellipse.  The inner
  ellipse corresponds to a disk of radius 2$\deg$, (27 kpc), with the
  same inclination as the main M31 disk.  }
\label{map}
\begin{flushleft}

\end{flushleft}
\end{center}
\end{figure}

\section{The bulge, discs and halo of M31}

The first recorded observation of M31 was made in 964 AD in 
the `Book of constellations and fixed stars' written by the Persian astronomer, 
Abd al-Rahman al-Sufi, who described it as `a small cloud' in the night sky. 
In the centuries that have passed since, M31 has been a 
popular target for astronomers, and much has been learned 
about its structure. M31 is a spiral galaxy of SA(s)b type, 
with a significant bulge, a classical thin stellar disc, 
a vast extended stellar disc and a metal poor halo. In 
this section, we outline the properties of each of these 
components. 

First we discuss the bulge component. Numerous studies have show that
M31 possesses a classical bulge, with a Sersic index of $\sim2$ and an
effective radius of 1.93 kpc \citep{kormendy99,seigar08}. It is
largely supported by random motions, although recent work by
\citet{saglia10} has found evidence for rotation in the innermost
regions. Saglia et al.  also find the bulge to be dominated by an old
stellar population (age $\gta 12$~Gyrs) of roughly solar metallicity,
with a large velocity dispersion of 166-170$\kms$. This component is
dominant out to about $8'$ ($\sim2\kpc$), at which point the disc
begins to dominates the surface brightness profile of the galaxy
\citep{saglia10}, however according to \citet{merrett06} the bulge can
be traced out as far as 10 effective radii, equivalent to $\sim15\kpc$
meaning that some of our innermost disc fields may be subjected to
minor contamination from this component.

Studies of the thin stellar disc of M31 have been performed by a
number of authors (e.g. \citealt{walterbos88,ferguson01}; I05;
\citealt{ibata07,mcconnachie09}), and have challenged our previous
notions of the structure of classical discs. With a scale length of
5.9 kpc (\citealt{walterbos88}, corrected for assumed distance to M31
of 785 kpc, and \citealt{merrett06}), it is more extensive than that
of our Galaxy, and also appears to be forming stars at a lower rate
\citep{walterbos94}.  And while it is a characteristic feature of the
surface brightness profiles of stellar discs to steeply decline at
3--4 scale lengths \citep{vanderkruit81,pohlen00}, which corresponds
to 18--24 kpc in M31, a spectroscopic study by \citet{ibata05}
uncovered a vast, extended disc component that can be traced out to
distances of $\sim40$~kpc ($\sim8$ scale lengths) from the centre of the
galaxy that has an exponential surface density profile that is very
similar to the inner disc. While this structure is rather clumpy, on
average it appears to follow on smoothly from the classical inner
disc, although perhaps with a slightly larger scale length of
6.6$\pm0.6\kpc$ and with a slight lag behind circular velocities at
large radii ( $\langle\Delta v\rangle=20\kms$, I05). It is dynamically
cold, with a velocity dispersion ranging from 20--40$\kms$, leading
I05 to conclude that it is likely not a thickened disc. Whether this
extended component is truly separate from the classical thin disc, or
merely an extension of it that shows some evidence of heating and
warping at larger radii where the disc is more sensitive to
perturbations from mergers and interactions, is unclear. Owing to the
similarity of the thin and extended discs, we will refer to them both
as the `thin disc' throughout this paper. Where we wish to make a
distinction between the two, we shall use the terminologies
`classical' and `extended' disc.

The presence of a smooth, pressure supported metal-poor halo in M31 eluded
detection until very recently. In 2006, two groups
\citep{chapman06,kalirai06} independently identified such a component
using the DEIMOS instrument on the Keck II telescope.  Centred on the
systemic velocity of M31, with a central velocity dispersion of
152$\kms$, and showing no strong evidence of rotation, both groups
found this component to be metal poor with an average metallicity of
$\feh=-1.4\pm0.2$ \citep{chapman06}. \citet{kalirai06} were able to
trace this component out as far as 165 kpc from the centre of the
galaxy although there is an inevitable confusion with the halo of M33
at these large distances \citep{ibata07,koch08, mcconnachie09}.

The halo of M31 is also a known host to a number of kinematic
substructures, such as the GSS, the tangential streams that cross the
SE minor axis, the western shelf and a wealth of substructure in the
NE of the galaxy that is thought to be linked to the GSS. In the
following analysis, we will carefully consider the kinematics of all
these components to ensure any thick disc sample that we define is
free from contamination by any of these sources. We shall discuss this
in greater detail in \S3.

\section{Observations and field selection}

\begin{figure}
\begin{center}
\includegraphics[angle=0,width=0.8\hsize]{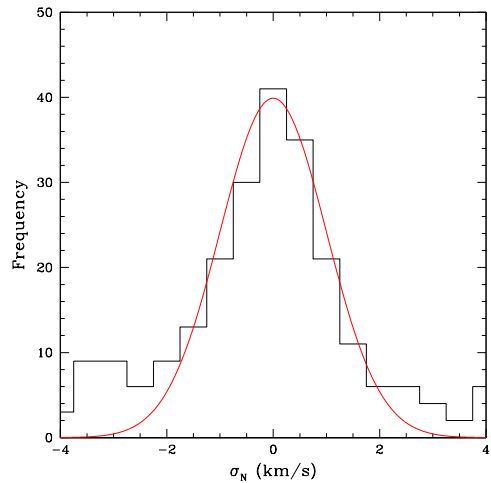}
\caption{A histogram showing the normalized error distribution for
  repeat measurements of the same stars in two of our fields that were
  observed with the 600 lines/mm grating. The normalized error,
  $\sigma_N$ incorporates the velocity differences between the repeat
  measurements (v$_1$ and v$_2$), and the Monte-Carlo uncertainties
  calculated for each observation ($\sigma_1$ and $\sigma_2$). In
  order to reproduce a unit Gaussian distribution for our
  uncertainties, we also include an additional error term, $\epsilon$,
  which accounts for any systematic uncertainties we have not
  included. We find $\epsilon$=5.6 kms$^{-1}$ for this setup. For
  fields using the 1200 lines/mm setup, we using the \citet{simon07}
  value of $\epsilon=2.2\kms$.}
\label{errors}
\end{center}
\end{figure}

A detailed description of the observational methodology and target
selection employed in the survey is given in I05, which we briefly
summarize here. Using Colour-Magnitude Diagrams (CMDs) from both the
Canada France Hawaii (CFHT) and Isaac Newton (INT) telescopes
\citep{ferguson02,ibata07,mcconnachie09}, we selected targets for
observation by prioritising Red Giant Branch (RGB) stars in M31 with
$20.5\leq i\leq21.2$ and colours $1.0\leq (V-i)_0\leq4.0$ (priority
A), then filling the remainder of the masks with stars with $I\le22.0$
that are unsaturated (priority B), where the V and I colours are
transformed from their native $g$ and $i$ colours using the relations
described in \citet{mcconnachie04b} and \citet{ibata07}.  We used a
combination of standard DEIMOS multislit mode for low density fields,
such as the halo, and our own minislitlet approach which allowed us to
target $>600$ stars per mask in more crowded regions, such as the
disc. Our observational setup covers the range of the Calcium Triplet
(Ca II) lines at 8498, 8542 and 8662\AA, a prominent absorption
feature that can be used both to measure radial velocities, and as a
metallicity indicator. To obtain velocities, we cross-correlate all
observed stars with a template Ca II spectrum. We estimate the errors
on our velocities by following the procedures of \citet{simon07} and
\citet{kalirai10}. First, we make an estimate of our velocity
uncertainties for each observed star using a Monte Carlo method,
whereby noise is randomly added to each pixel in the spectrum,
assuming a Poisson distribution for the noise and the velocity is
recalculated using the same cross correlation technique described
above. This procedure is repeated 1000 times, and then the error is
calculated to be the square root of the variance of the resulting mean
velocity. We combine this error with a systematic error, $\epsilon$,
which contains information on any errors we may not have accounted for
(for example, wavelength calibration error, misalignment of the mask
etc.). For the fields observed with the 600 line/mm grating, we
evaluate this error directly by using repeat measurements in fields
231Dis and 232Dis, a total of 332 stars. We define the normalised
error, $\sigma_N$ as:

\begin{equation}
\sigma_N=\frac{v_1-v_2}{(\sigma_1^2+\sigma_2^2+2\epsilon^2)^{1/2}}
\end{equation}

\noindent where $v_1$ and $v_2$, $\sigma_1$ and $\sigma_2$ are the
velocities and errors of each measurement pair, and $\epsilon$ is the
additional random error required in order to reproduce a unit Gaussian
distribution with our data (shown in Fig.~\ref{errors}) This gives us
a systematic error for this setup of $\epsilon=5.6\kms$, slightly
lower than the value of $\epsilon=6.2\kms$ derived by
\citet{collins10} for the same setup, though we note that their
measurement was based on repeat observations of 47 stars, compared
with our much larger data set of 332 stars. The typical uncertainties
for these measurements above a threshold of S:N = 3, are 5-10$\kms$.

The Ca II features also provide us with a method for measuring the
spectroscopic metallicity of our observed sample. Following the
procedure of \citet{rutledge97} and \citep{battaglia08}, we fit
Gaussian functions to the three Ca II peaks to estimate their
equivalent widths (EWs), and calculate [Fe/H] using equation (1)

\begin{equation}
[Fe/H]=-2.66+0.42[\Sigma Ca+0.64(V_{RGB}-V_{HB})]
\end{equation}

\noindent where $\Sigma$Ca=0.5EW$_{8498}$+EW$_{8542}$+0.6EW$_{8662}$,
$V_{RGB}$ is the magnitude (or, if using a composite spectrum, the
average magnitude) of the RGB star, and $V_{HB}$ is the mean
$V$-magnitude of the horizontal branch (HB). Using $V_{HB}-V_{RGB}$
removes any strong dependence on distance or reddening in our
calculated value of [Fe/H], and gives us the Ca II line strength at
the level of the HB. For M31, we set this value to be 25.17
\citep{holland96}. We note that this assumed value is sensitive to age
and metallicity effects, see \citealt{chen09} for a discussion,
however owing to the large distance of M31, small differences in this
value within the disc of M31 will have a negligable effect on
metallicity calculations. For individual stars, these measurements
carry large errors ($\gta0.4$ dex), but the errors are significantly
reduced when stacking the spectra into a composite in order to measure
an average metallicity for a given population.

\subsection{Field selection and sample definition}

\begin{table*}
\begin{center}
\caption{Properties of fields analysed in this work}
\label{fprops}
\begin{tabular}{lcccccccc}
\hline
Field  & Date observed & $\alpha_{J2000}(hh:mm:ss)$  & $\delta_{J2000}(\deg:':'')$ & Grating & P.A. & Exp. time (s) & No. targets & R$_{proj}$ (kpc)\\
\hline
228Dis & 23/09/2006 & 00:40:50.56 & 40:43:54.0 & 1200 lines/mm & 90$\deg$ & 3600s & 301 & 9.8\\
227Dis & 23/09/2006 & 00:39:37.40 & 40:50:42.0 & 1200 lines/mm & 90$\deg$ & 3600s & 312 & 15.6\\
166Dis & 03/10/2005 & 00:39:17.89 & 40:42:18.0 & 1200 lines/mm & 90$\deg$ & 3600s & 209 & 15.8\\
106Dis & 30/08/2005 & 00:39:10.00 & 40:39:00.0 & 1200 lines/mm & 90$\deg$ & 3600s & 257 & 16.1\\
105Dis & 30/08/2005 & 00:39:00.00 & 40:28:12.0 & 1200 lines/mm & 90$\deg$ & 3600s & 271 & 16.2\\
224Dis & 25/09/2006 & 00:38:50.00 & 40:20:30.0 & 1200 lines/mm & 90$\deg$ & 3600s & 265 & 16.6\\
232Dis & 05/10/2006 & 00:38:50.00 & 40:20:00.0 & 600 lines/mm  & 90$\deg$ & 16200s& 184 & 16.6\\
104Dis & 30/08/2005 & 00:38:50.00 & 40:20:00.0 & 1200 lines/mm & 90$\deg$ & 3600s & 271 & 16.7\\
220Dis & 22/09/2006 & 00:38:00.00 & 40:06:12.0 & 1200 lines/mm & 90$\deg$ & 3600s & 322 & 20.3\\
213Dis & 22/09/2006 & 00:38:11.60 & 40:06:12.0 & 1200 lines/mm & 90$\deg$ & 3600s & 155 & 20.5\\
102Dis & 30/08/2005 & 00:38:00.00 & 40:00:00.0 & 1200 lines/mm & 90$\deg$ & 3600s & 268 & 21.5\\
231Dis & 05/10/2006 & 00:38:00.00 & 40:00:00.0 & 600 lines/mm  & 90$\deg$ & 16200s & 185 & 21.6\\
223Dis & 25/09/2006 & 00:37:12.00 & 39:57:00.0 & 1200 lines/mm & 90$\deg$ & 3600s & 304 & 23.2\\
101Dis & 30/08/2005 & 00:38:00.00 & 39:54:00.0 & 1200 lines/mm & 90$\deg$ & 3600s & 275 & 23.5\\
222Dis & 22/09/2006 & 00:37:12.49 & 39:48:06.0 & 1200 lines/mm & 90$\deg$ & 3600s & 298 & 24.9\\
221Dis & 22/09/2006 & 00:37:11.97 & 39:45:00.0 & 1200 lines/mm & 90$\deg$ & 3600s & 303 & 25.8\\
50Disk & 16/09/2004 & 00:37:35.29 & 39:33:55.0 & 1200 lines/mm & 90$\deg$ & 3600s & 216 & 30.1\\
107Ext & 30/08/2005 & 00:35:28.00 & 39:36:19.1 & 1200 lines/mm & 90$\deg$ & 3600s & 265 &31.0\\
w11old & 30/09/2002 & 00:35:27.02 & 39:37:15.3 & 1200 lines/mm & 90$\deg$ & 3600s & 95  &31.0\\
167Hal & 03/10/2005 & 00:34:30.24 & 39:23:58.7 & 1200 lines/mm & 90$\deg$ & 3600s & 205 &34.2\\
148Ext & 04/10/2005 & 00:37:07.23 & 39:12:00.0 & 1200 lines/mm & 90$\deg$ & 3600s & 211 & 39.6\\
\hline
\end{tabular}
\end{center}
\end{table*}

\begin{figure}
\begin{center}
\includegraphics[angle=0,width=0.99\hsize]{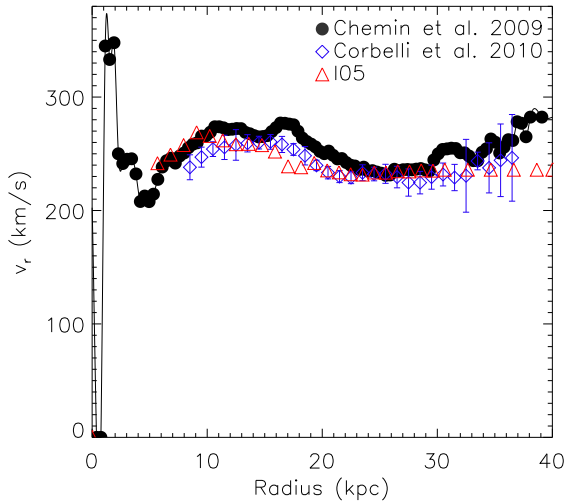}
\caption{Here we show a number of HI rotation curves for
  M31. Throughout this work, we use results based on the
  \citet{chemin09} work, shown as filled black circles. We also show
  rotation curves from \citet{corbelli10} and I05, which differ
  slightly from the Chemin et al. 2009 curve in the outermost
  regions. We find that using these curves vs. the \citet{chemin09}
  results do not affect our results.}
\label{rotcurve}
\end{center}
\end{figure}

\begin{figure}
\begin{center}
\includegraphics[angle=0,width=0.99\hsize]{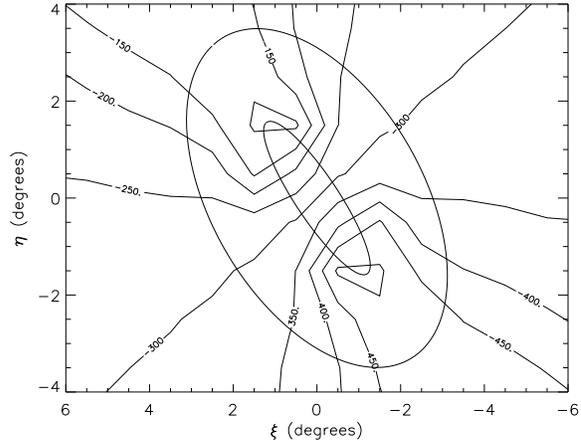}
\caption{A contour map of the expected velocities of stars in circular
  orbits in the disc of M31. This was constructed using our simple
  model as discussed in \S3.  }
\label{velmap}
\end{center}
\end{figure}

In order to detect a thick disc component in M31 kinematically, we
need to measure the velocities of stars within our sample relative to
some model for the velocity of stars within the thin disc of the
galaxy. If a thick disc is present, we should observe a population
that lags behind the thin disc in terms of its rotational
velocity. The component is also expected to have a larger velocity
dispersion than the thin disc.  This is observed in the MW, where the
thick disc lags the thin by between 20-50 kms$^{-1}$
\citep{chiba00,soubiran03} and has an average rotational velocity
dispersion of $\sigma_{V_{\phi}}$=57 kms$^{-1}$ \citep{carollo10}. For
the purposes of this work, we shall use an updated version to the disc
model of I05. In this model, we assume circular orbits for all stars
about the centre of M31 and interpolate their velocities from the HI
rotation curve of \citet{chemin09}, which is shown in
Fig.~\ref{rotcurve} as the solid black points. This rotation curve
differs from that adopted by I05, particularly in the
outermost regions. They used a compilation of CO data
from \citet{klypin02} and HI data from \citet{brinks84}, which we also
show in Fig.~\ref{rotcurve} as red triangles. We also show the HI rotation curve derived by
\citet{corbelli10} from a WRST survey \citep{braun09} as blue
squares. Using either of these rotation curves as opposed to that of
Chemin et al. leads to differences in our interpolated velocities of
order a few--$20\kms$, however there is a negligible effect on the
dispersions within particular populations, and so the adoption of any
of these curves would give us consistent results when analysing the
global properties of the stellar discs in this work. We assume an
inclination for M31 of 77$\deg$ \citep{walterbos88}, and adopt
parameters for the thickness of the disc identical to those used in
I05, with a constant thickness for the disc of 350 pc (which is
roughly the thickness of the MW disc, \citet{ivezic08}) out to 16 kpc,
at which point we assume the stellar disc begins to flare with a scale
height that increases linearly with radius. We set a maximum scale
height of 1.69 kpc at radius of 30.5 kpc, beyond which we assume that
the disc has constant thickness. We then integrate along the line of
sight through this flaring exponential disc and project the velocities
of objects on circular orbits about M31 onto the line of sight. This
produces an average velocity map for the disc of M31, which we display
in Fig.~\ref{velmap}.

Once our disc model has been constructed, we need to select
a sample of fields from our DEIMOS survey that will
provide the most reliable kinematic comparison with respect to the
velocity of the disc. As the disc of M31 is not infinitesimally thin,
but possesses some unknown scale height, any line of sight taken
through the galaxy traverses a significant depth. Given the
inclination of M31, some lines of sight will traverse larger depths
than others, which could have the effect of smearing out the
velocities of objects with respect to the disc model. This is
illustrated in Fig.~8 of I05. They find that objects along the major
axis of M31 are less susceptible to this effect than those that are
located off the major axis, and therefore we limit our initial study to
fields along the major axis.

A further complication in field selection arises from MW
contamination. Our colour selection criteria means that we inevitably
observe Galactic K dwarf stars within our sample, as these lie
coincident with M31 RGB stars in the CFHT and INT CMDs. Eliminating
these stars from our sample is straightforward in the South West (SW)
region of M31, as the disc of Andromeda and the halo of the MW occupy
distinct positions in heliocentric velocity space. Assuming the
Besancon model is a good description of the foreground populations in
the direction of M31, it can be shown that the Galactic population
peaks at v$_{hel}$=-61kms$^{-1}$, and the contribution of MW K dwarfs
to our sample at v$_{hel}\leq-100$kms$^{-1}$ is very low
(\citealt{robin04}, I05). Given that the average rotational velocity
in the SW of M31 is less than -300 kms$^{-1}$ (I05), we are able to
cleanly separate M31 stars from MW field stars. However, in the North
East (NE) of M31, the average heliocentric velocity of the M31 disc
typically ranges between -100 kms$^{-1}$ and -200 kms$^{-1}$,
resulting in a significant overlap between Galactic and M31
populations, making it difficult to distinguish between the two. While
it is possible to remove some of this contamination by examining the
strength of the Sodium doublet (NaI), located at a rest wavelength of
$\sim8190\Aa$, this is not a perfect discriminator. One can also
eliminate some foreground contamination via a comparison of
photometric and spectroscopic metallicities \citep{gilbert06}, but
given the uncertainties on the individual spectroscopic [Fe/H] of our
observed stars (discussed above), we still retain a significant
population of contaminants within our sample. There is also
contamination in the NE from M31 substructure
(I05,\citealt{chapman06,richardson08}) which can be difficult to
separate from the M31 disc in the NE. For these reasons, we limit our
initial study to the SW major axis. We hope to analyse the NE
population in a future paper. These criteria leave us with a sample of
21 fields along the SW major axis, highlighted in red in
Fig.~\ref{map}. The positions and properties of these fields can be
found in Table~\ref{fprops}. Two of these fields (231Dis and 232Dis)
were observed as part of our ultra-deep M31 disc survey (Chapman et
al. in prep.) and were integrated for 4.5 hours with the 600 line/mm
grating, which allowed us to make more robust measurements of
individual stellar metallicities than for our other fields.

\begin{figure}
\begin{center}
\includegraphics[angle=0,width=0.99\hsize]{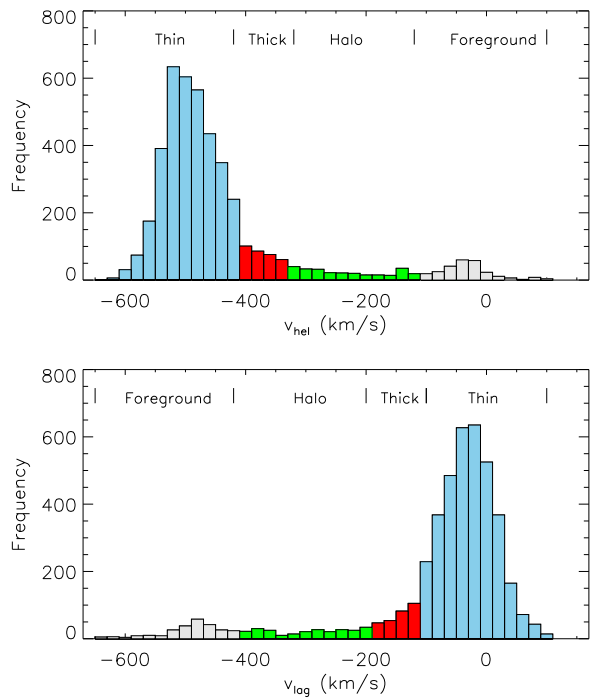}
\caption{Histograms for both heliocentric (top) and disc lag (bottom)
  velocities of all stars within our sample of 21 fields. Regions
  expected to be inhabited by thin disc (light blue), thick disc
  (red), halo (green) and MW foreground (grey) are highlighted. }
\label{contam}
\end{center}
\end{figure}

\begin{figure*}
\begin{center}
\includegraphics[angle=0,width=0.99\hsize]{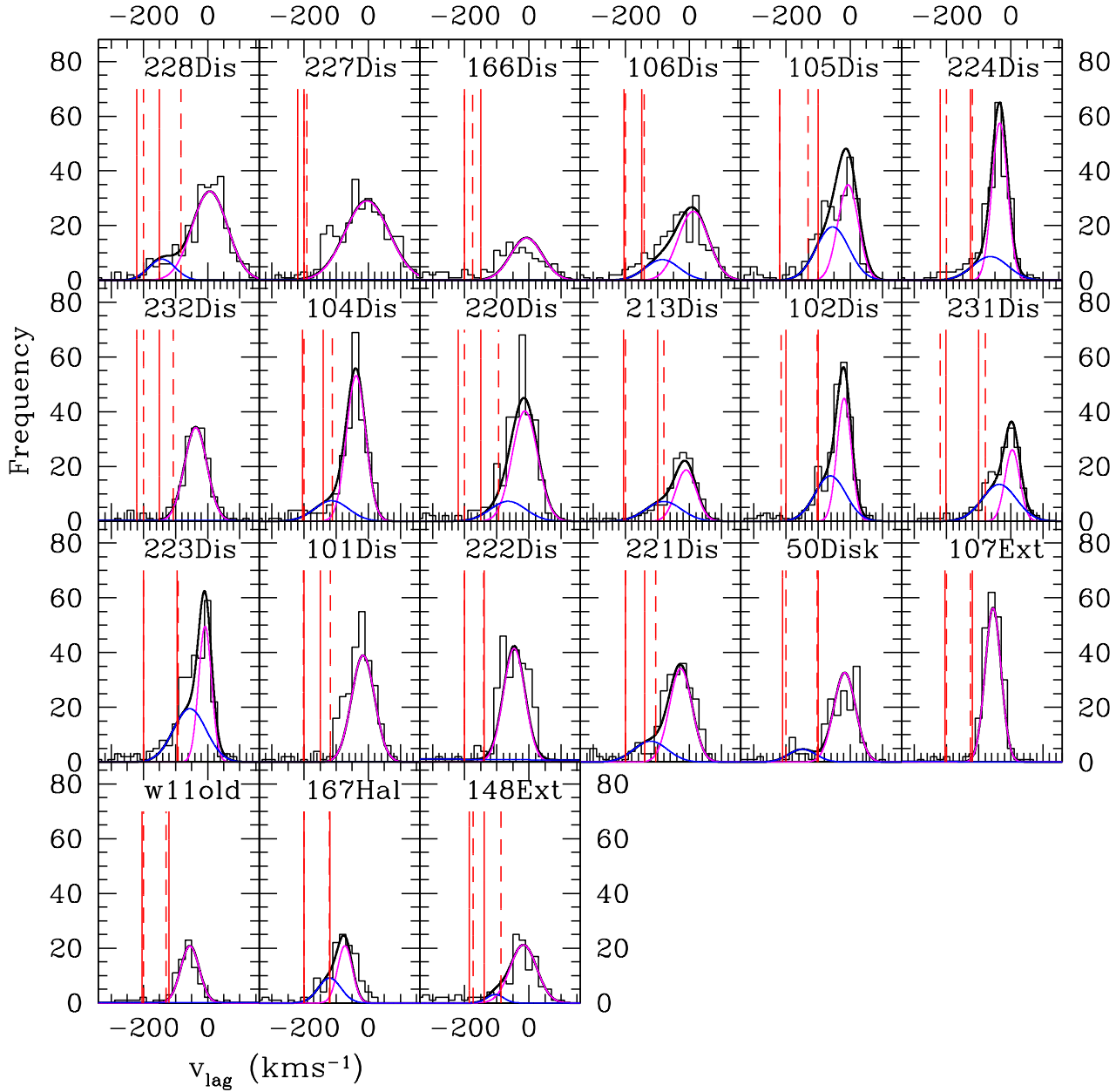}
\caption{Our initial sample of fields, selected for their position
  along the South West major axis as described in the text, are shown
  in order of increasing (projected) radius. Gaussian fits
  indicating the thin and (where applicable) thick disc are shown as
  magenta and blue curves respectively. Our selection criteria are
  overlaid, with the dashed lines representing our 2$\sigma$ cuts and
  the solid lines representing our Gaussian cuts both of which sample
  roughly the same region of velocity space.  }
\label{fields}
\end{center}
\end{figure*} 

To isolate a potential thick disc population in our sample of 21
fields, we must first define a statistical measure of what constitutes
a lagging population with respect to the thin and extended discs. We
must also ensure that our definition is able to distinguish between
this population and that of the metal-poor M31 halo which, as it is a
non-rotating component \citep{chapman06,kalirai06}, also lags behind
the disc. In Fig.~\ref{contam}, we display two histograms, one with
the heliocentric velocities ($v_{hel}$) of the stars in our 21 fields,
and one with their velocities with respect to the disc ($v_{lag}$) and
we highlight the regions we expect each of these components to
inhabit, along with where we expect to see contamination from halo
K-dwarfs in the MW.

\begin{figure}
\begin{center}
\includegraphics[angle=0,width=0.99\hsize]{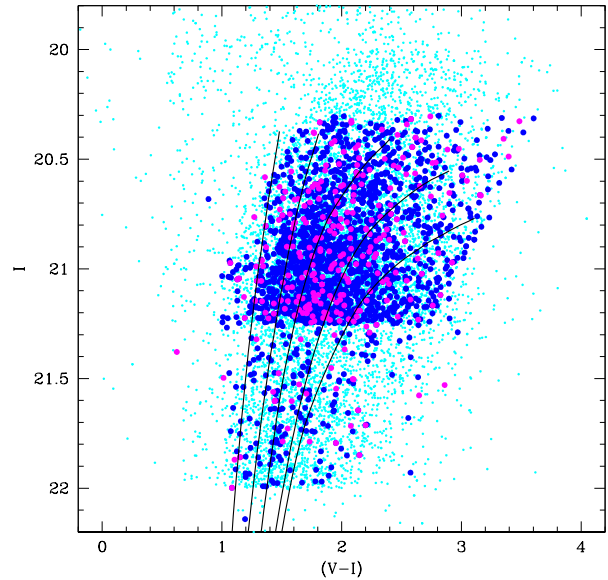}
\caption{CMD for our thin (blue points) and thick (magenta points)
  populations in standard Landolt V and I colours. Dartmouth
  isochrones with $\afe=+0.2$ and an age of 8 Gyrs are overlaid with
  metallicities from left to right of [Fe/H]=-1.5, -1.0, -0.5,
  -0.4. The colours of the remaining stars in our DEIMOS survey are
  also plotted in light blue.}
\label{cmd}
\end{center}
\end{figure}

 We do this using two separate methods. The first is to fit Gaussians
 to both a disc component, located on or around
 v$_{lag}$=0~kms$^{-1}$, and a broad halo component centered on or
 around v$_{lag}$=-300~kms$^{-1}$. We then define a thick disc
 population to encapsulate anything that lags the thin disc by
 $>2\sigma$ of the thin disc peak value and we implement a lower cut
 on this population by requiring the contribution from the halo would
 be $<$1 star per velocity bin (20 kms$^{-1}$), thus minimizing the
 contamination. In fields where there is no obvious halo component to
 fit to, we use a Gaussian centered on -300 kms$^{-1}$ with a
 dispersion of 90~kms$^{-1}$ \citep{chapman06}, and normalize it with
 respect to the thin disk by assuming that the halo contributes
 $\sim$10\% to the total number of stars within the field (a
 conservative estimate, given that the stellar halo contributes
 $<<10$\% to the total stellar light in disc galaxies). The second
 method is to fit multiple component Gaussians to each of the
 fields. We apply a Gaussian Mixture Modelling (GMM) technique, which
 allows the number of Gaussians to vary freely between 1 and
 7 components. To discern which model best fits the data, we apply a
 likelihood ratio test (LRT) to the resulting probabilities of the
 fits.  The use of the LRT in astronomy was popularised by
 \citet{cash79}, and is often used in the literature to determine
 whether the properties of an observed stellar population can be well
 described by single vs. multiple Gaussian components
 (e.g. \citealt{ashman94,carollo10}). The LRT compares the likelihoods
 of nested models (in our case, a mixture of Gaussian components) to
 determine whether applying a model with additional parameters
 produces a significantly better fit than a simpler model. This is
 done by calculating the LRT statistic, $-2\rm ln( \mathcal{L}_1/
 \mathcal{L}_2)$, where $ \mathcal{L}_1$ and $ \mathcal{L}_2$
 represent the likelihoods of the simple and complex model
 respectively, and comparing it with a $\chi^2$ distribution with
 degrees of freedom equal to the difference between the number of
 parameters in the two models (3 in our case). For a model with
 additional parameters to be accepted as a statistically better fit,
 this ratio must be greater than 7.82 which corresponds to a P-value
 of $<0.05$. In general, this technique converges on fits with three
 components (a thin disc, a halo and a thick population) though there
 are a few exceptions. We shall discuss these fits in greater detail
 in the following section. Where this technique converges on fits that
 identify a lagging component that is distinct from both the thin disc
 and halo, we define a sample of highly probable thick disc stars by
 applying a standard Bayesian classification scheme to assign each
 star a probability of being a member of the thin disc $P$(thin),
 thick disc, $P$(thick), or halo, $P$(halo), population based on their
 velocity, and the properties of the Gaussian fits to each population
 on a field by field basis. We define a star as being a highly
 probable member of the thick disc if $P$(thick)$\geq0.997$. The
 results of both these techniques can be seen in
 Fig.~\ref{fields}. The velocity cuts for stars selected using our
 2$\sigma$ technique are shown as dashed lines, and the range of
 velocities selected using the Bayesian classification technique are
 marked with solid lines. It can be seen that both techniques isolate
 a very similar population. In Fig.~\ref{cmd}, we plot a CMD showing
 the V-I colours of the thin (blue points) and thick (magenta points)
 populations, and we overlay Dartmouth isochrones \citep{dart08} of
 $\afe=+0.2$ and an age of 8 Gyrs (in line with the estimated range of
 ages for the thin disc of 4--8 Gyrs, \citealt{brown06}) with
 metallicities ranging from $\feh=-0.4$ to $\feh= -1.5$. Both thin and
 thick populations inhabit roughly the same region in this CMD, which
 we shall discuss in more detail in \S4.3.

Finally, we note that in both selection methods, we expect some cross
contamination between the thin and thick disc components, as the two
populations significantly overlap. However, we assume this
contamination will be lower in our cuts based on the Gaussian fits, as
these are more conservative. Therefore, we use these cuts
predominantly in this paper when referring to clean thin and thick
disc samples. We have also assumed both components have symmetric,
Gaussian distributions in velocity, which may not be the case, and
this could cause further contamination if the populations are
skewed. We also expect some contamination from the halo, however,
given that the disc is the dominant population in all our fields, we
expect this contamination to be negligible in comparison to the cross
contamination between the discs.

\subsection{Testing the statistical significance of our sample}

Before we analyse our sample, we test the significance of our thick
population to ensure it is not merely consistent with noise above a
thin disc population plus smooth halo component. To do this, we fit a
single Gaussian to the disc and halo components, as described above,
then calculate the deviation of the data from the fit for all
velocities greater than the peak disc velocity (i.e. the right hand
side of the disc fit), normalizing it to the expected contribution
from the Gaussian in this region. We then define the noise to be 1.5
times the median absolute deviation of this sample. We repeat this
exercise for all velocities less than the thin disc peak and greater
than -200~$\kms$ in the lag frame, in this case comparing to the
expected contribution from both thin disc and halo fits. This allows
us to work out the significance of our thick disc population,
$\sigma_{conf}$. In all cases where the GMM identified a thick disc
component, we find that our excess above the thin disc plus halo model
has a significance of $>3\sigma$ (see table~\ref{kprops}). For the
fields where the GMM converged on a 2 Gaussian fit (232Dis, 222Dis,
107Ext and w11old), we find $\sigma_{conf}<3\sigma$. We also identify an
additional 3 fields (101Dis, 166Dis and 227Dis), where
$\sigma_{conf}<3\sigma$. Two of these fields are located at radii of
$\sim15\kpc$, where there may be residual contamination from the bulge
component. This may also explain the large dispersions (of order
50$\kms$) seen in our innermost fields. Excluding these, we are left
with 14 of our 21 ($2/3$) fields where we confidently detect a
thick component. We shall focus on these fields in the remainder of
our analysis, but we shall discuss the significance of the
non-detections in \S5.

\section{Results}

\subsection{Kinematic and structural properties of the thin and thick discs}

In this section we present measurements for the kinematic and
structural properties of the thin and thick discs. Properties of
individual fields can be found in Table~\ref{kprops}, while the
average properties for both components can be found in
Table~\ref{avprops}.

\subsubsection{Velocity lag and dispersion profiles}

\begin{figure}
\begin{center}
\includegraphics[angle=0,width=0.9\hsize]{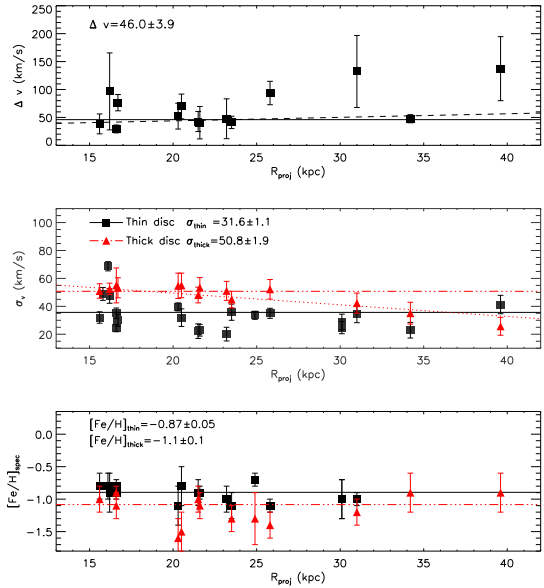}
\caption{{\bf Top panel:} The difference in velocity, $\Delta v$,
  between the thin disc and thick component as a function of projected
  radius. This lag appears to be approximately constant as a function
  of radius, with an average lag of $46.0\pm3.9\kms $. There appears
  to be a slight in crease in lag in the outermost part, however this
  is largely driven by fields that lie off the major axis of M31, and
  therefore the velocities are less reliable. {\bf Middle panel:
  }Dispersion, $\sigma_v$, of both thin (black squares, solid line)
  and thick (red triangles, dot-dashed line) components are plotted as
  a function of projected radius. The thin disc appears to maintain a
  constant dispersion of $\sigma_{thin}$=35.7$\pm1.0\kms$, however the
  thick component appears to decrease somewhat at larger radii. {\bf
    Bottom panel: }Average spectroscopic metallicity of thin and thick
  components as a function of projected radius. Neither component
  evolves with radius.}
\label{summary}
\end{center}
\end{figure}

\begin{figure}
\begin{center}
\includegraphics[angle=0,width=0.9\hsize]{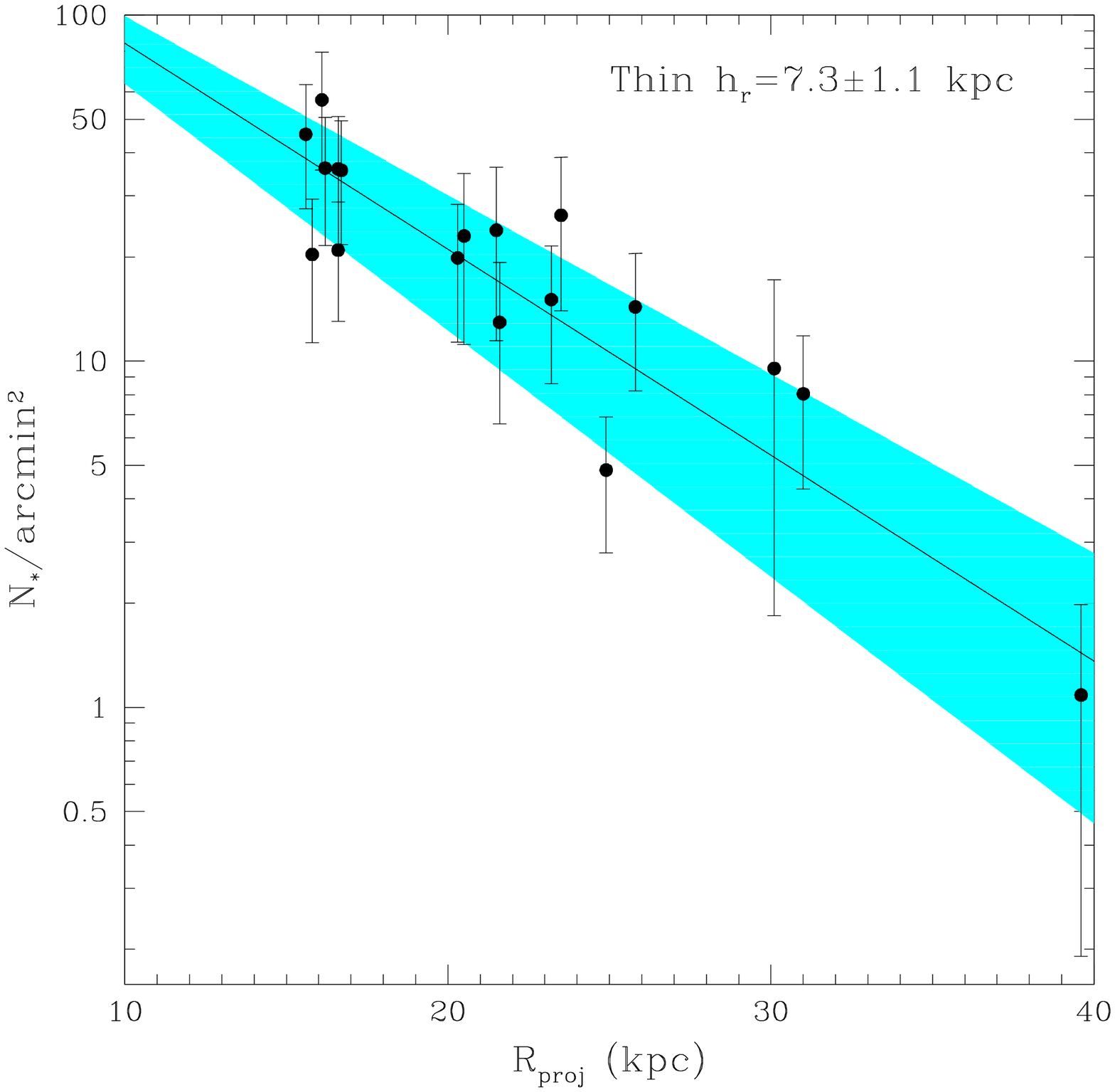}
\includegraphics[angle=0,width=0.9\hsize]{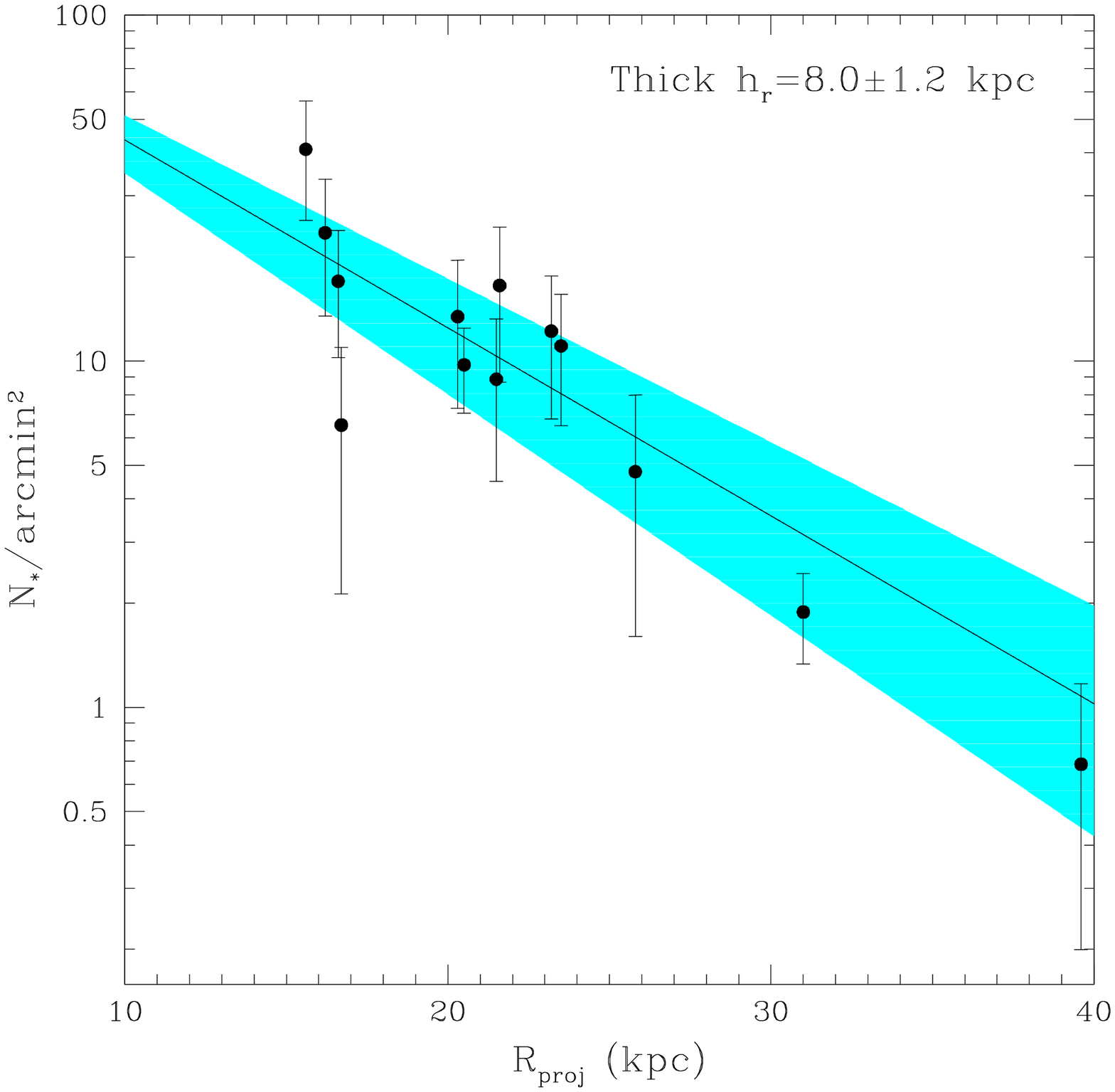}
\caption{{\bf Top panel:} Plot of the density of stars
  (N$_*$/arcmin$^2$) in the thin disc against R$_{proj}$. The
  densities are calculated by first separating our sample by their
  target prioritisation (A or B, see \S3), then counting all stars
  with $v_{thin}>v_{lag}>v_{thin}+2\sigma_{thin}$ and multiplying
  these values by 2 (i.e. assuming the distributions are symmetric)
  for both prioritisations. We then calculate
  $\rho_*==n_sn_t/n_o-n_b$, and combine these results from priority A
  and B. Fitting an exponential profile to these points we deduce
  $h_r=7.3\pm1.1$~kpc. Solid line represents the best fit to the data
  from a weighted-least-squares routine, and the shaded region
  indicates the 1$\sigma$ errors from the fit. {\bf Bottom panel:} As
  above, for the thick disc. Here we count all stars with
  $v_{thick}-2\sigma_{thick}>v_{lag}>v_{thick}$ and multiply by two
  again. Fitting an exponential profile to these points we deduce
  $h_r=8.0\pm1.2$~kpc.}
\label{scale}
\end{center}
\end{figure}

\begin{figure*}
\begin{center}
\includegraphics[angle=0,width=0.3\hsize]{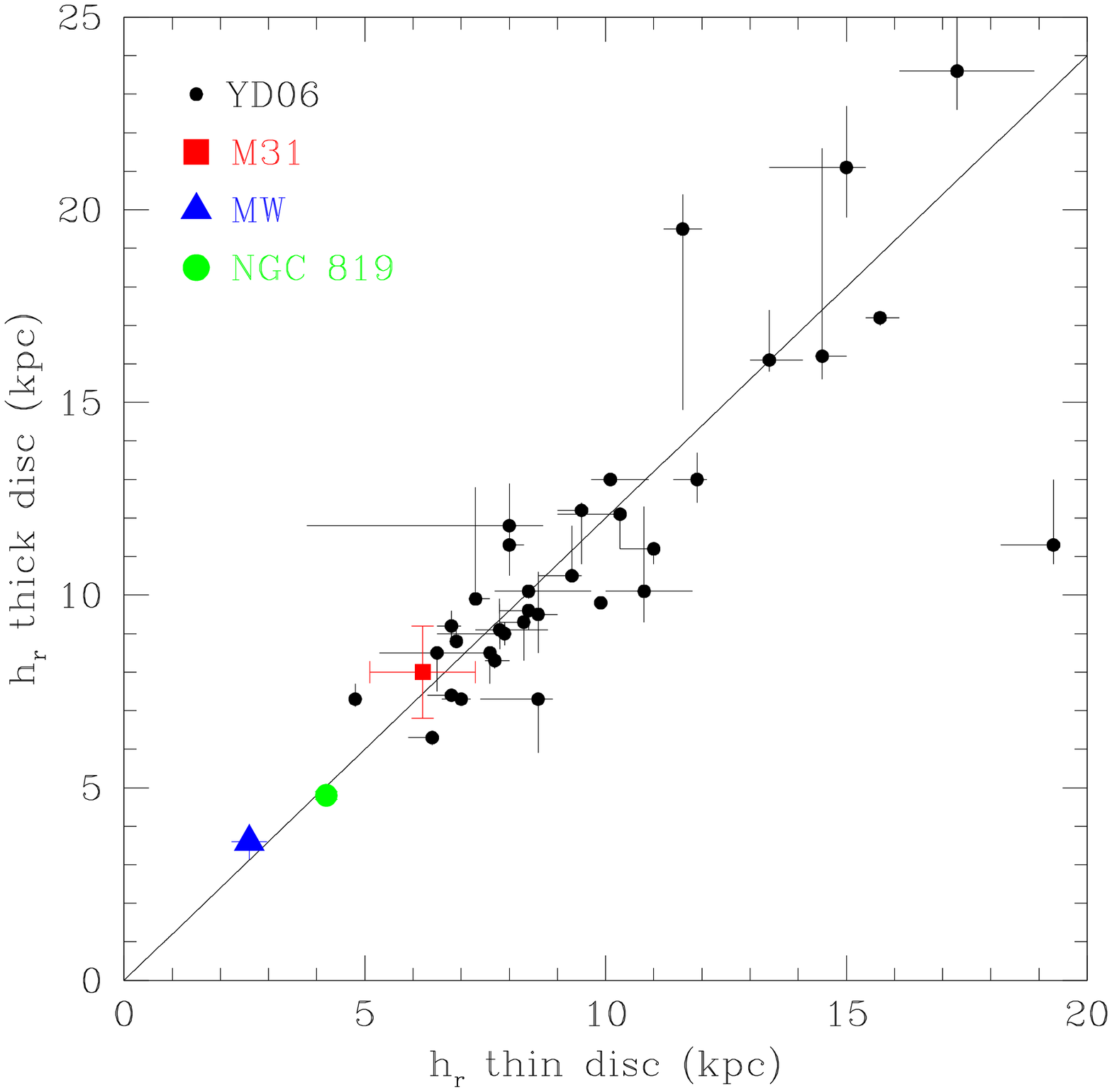}
\includegraphics[angle=0,width=0.3\hsize]{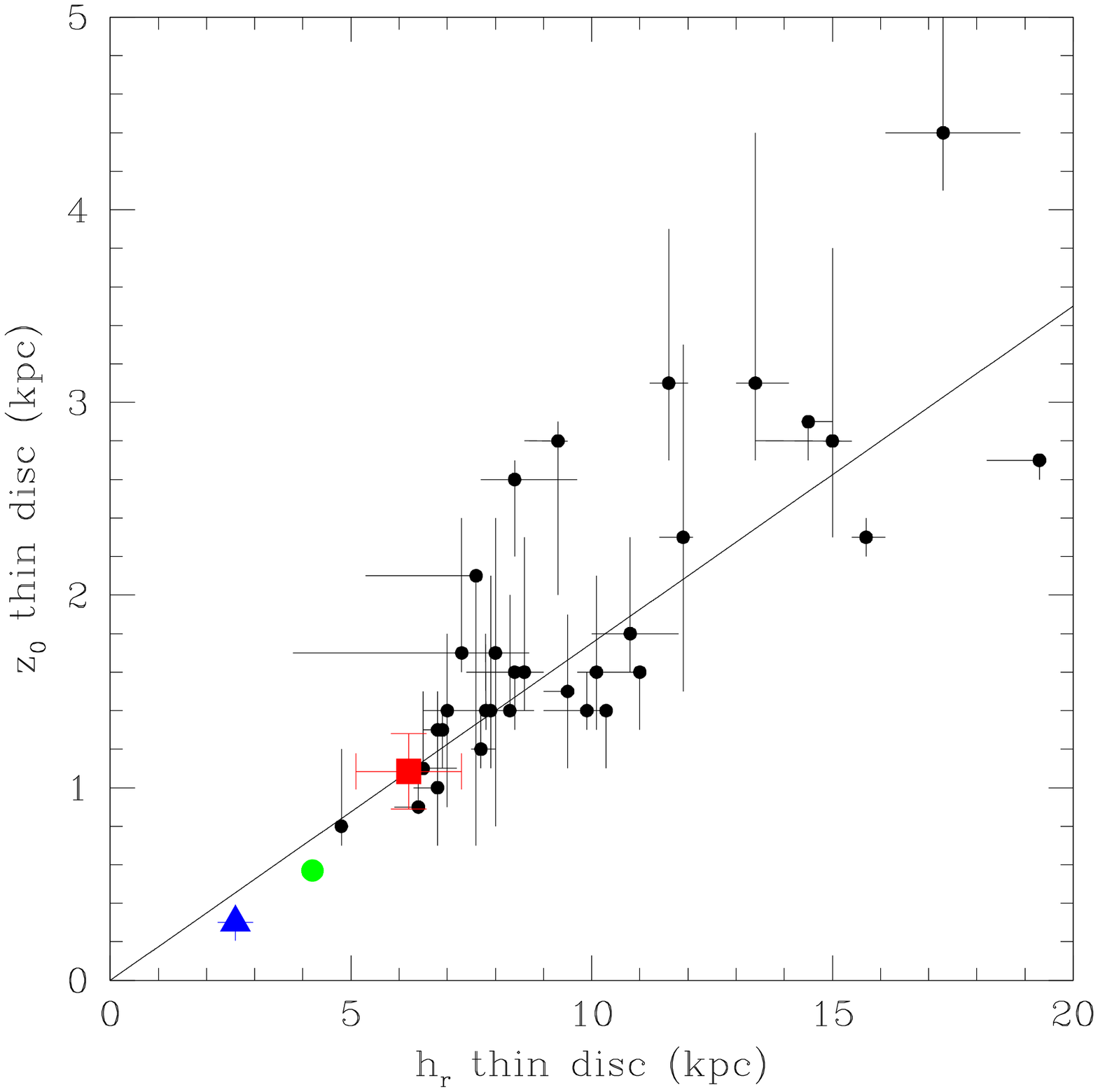}
\includegraphics[angle=0,width=0.3\hsize]{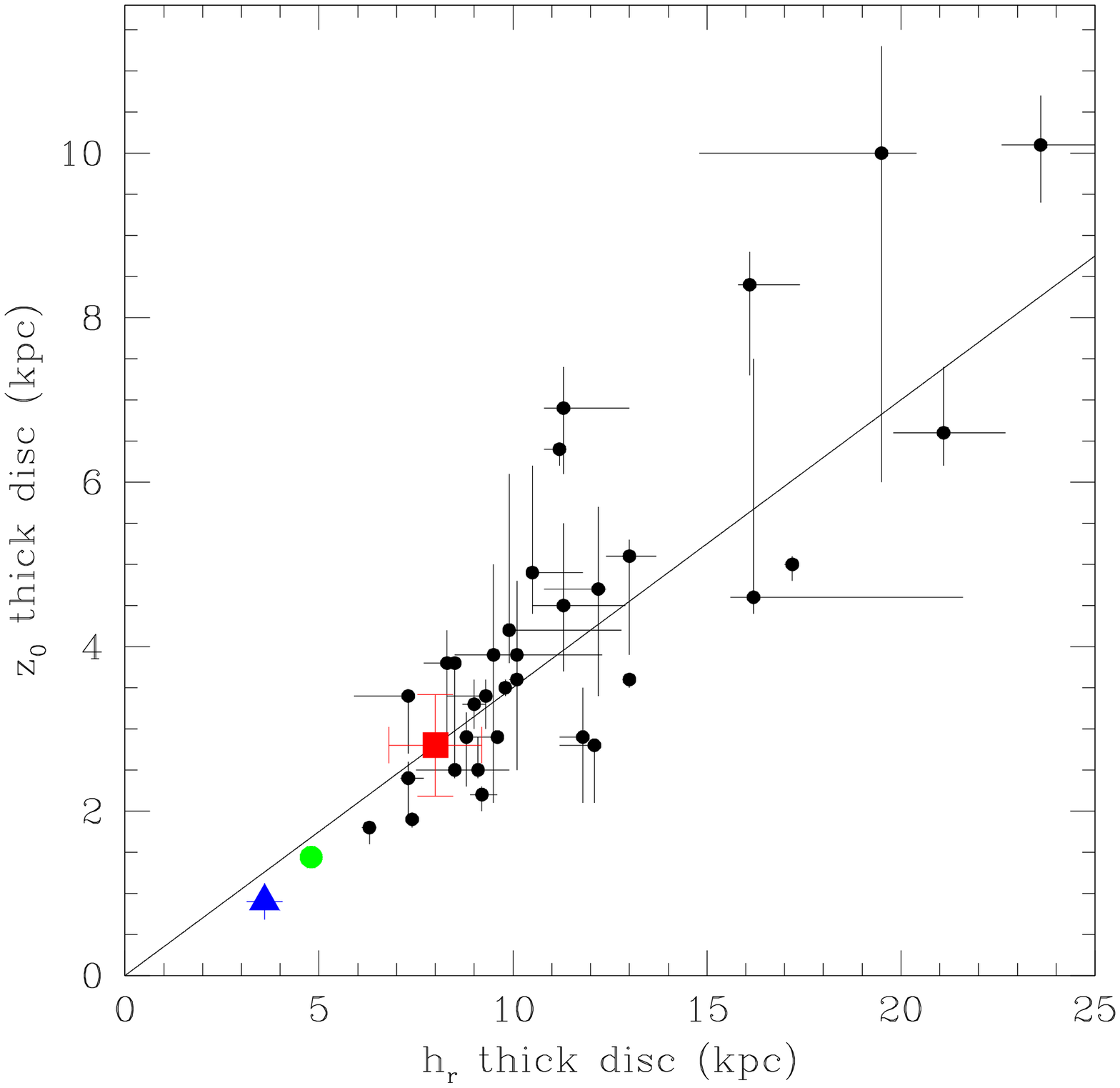}
\caption{Here we compare the scale lengths of the thin and thick discs
  in M31 to those of other galaxies with observed thick discs in order
  to infer their scale heights. {\bf Left panel: }Here we plot thin disc
  scale lengths against thick disc scale lengths for 34 external
  galaxies (YD06) as black points, and overplot the same measurements
  for the MW (blue triangle). We fit a linear relation to these points
  with a gradient of 1.3. We plot our result for M31 as a red square,
  and it is in excellent agreement with this relation. {\bf Centre panel:
  }Here we plot scale height, $z_0$, against scale height for the thin
  disc of the YD06 sample plus the MW and derive a best-fit linear
  function with a gradient of 0.17. From this, we can estimate a scale
  height for the M31 thin disc of 1.1$\pm0.2$~kpc, which we overplot
  on the relation as a red square. {\bf Right panel: } We now plot the
  same, but for the thick disc and find a best fit gradient of 0.35,
  and therefore infer a scale height for the thick disc in M31 of
  2.8$\pm0.6$~kpc, which is overplotted as a red square.}
\label{height}
\end{center}
\end{figure*}

In this first section, we initially address the thin and extended
discs of M31. In I05, the extended disc was identified as a stellar
disc that, while appearing in many respects to be similar to the
classical thin disc, was a separate entity that was clumpy in terms of
its structure, and lagged behind the classical disc in terms of its
kinematics. As we are limiting our study to one slice down the major
axis of M31, we do not attempt to comment on the global `clumpiness'
of this extended disc, but we return to the issue of the velocity lag
and distinction from the thin stellar disc. As we have analysed the
disc frame velocities for all our fields using a rotation curve that
differs from the one used in I05, it is useful for us to determine
whether the increasing lag with respect to the classical thin disc is
seen here also. In I05, they split their sample of 21 fields into an
inner (with R$_{proj}<20\kpc$) and outer (with 20$<$R$_{proj}<30\kpc$)
sample to determine the average properties of the disc and extended
disc. For their inner (classical) disc sample, they calculated an
average velocity for the disc in the disc lag frame of
$v_{lag}=-17.0\kms$ and a dispersion of $\sigma_v=50.0\kms$. In the
outer (extended) sample they calculated an average velocity of
$v_{lag}=-16.0\kms$ and a dispersion of $\sigma_v=51.0\kms$. If we
perform the same analysis for our study, we find an average lag of
$v_{lag}=-14.8\kms$ for our inner fields and $v_{lag}=-25.5\kms$ for
our outer fields. However, we note that this value is calculated with
the inclusion of fields 107Ext, w11old and 167Hal, which have very large lags of
$v_{lag}<-55\kms$ compared with the other fields. We note that these 
fields are located slightly off the semi major axis (see Fig.~\ref{map}) 
where our interpolated disc-frame velocities are subject to larger uncertainties. If
we exclude these fields, we find an average lag of $v_{lag}=-14.9\kms$,
very similar to our inner sample. We therefore conclude that there is
a negligible difference in the lags of the classical and extended disc
behind circular velocities. For these samples we also calculate
average dispersions of $\sigma_v=42.7\kms$ and $\sigma_v=30.0\kms$,
implying that the extended disc has a lower dispersion than the
classical disc. However, in our inner sample, we are more likely to see 
residual contamination from the bulge and we also have a large
proportion of fields for which we could not cleanly isolate the thick
disc ($\sim40$\% cf. $\sim20$\% in the outer sample). These factors may 
cause us to overestimate the dispersion of the disc in these
regions. From these results, we therefore find no concrete reason to
assume that the extended disc is a separate component from the
classical disc and we treat these two components as one thin stellar
disc in the remainder of our analysis.

By using the information from our Gaussian fits to the thin and thick
components, we can comment on their global kinematic properties, and
discuss any variation of these properties with radius. In
Table~\ref{kprops} we show the peak velocities and velocity
dispersions of both thin and thick (where applicable) components in
each field, with associated errors from the GMM fits. Where both thin
and thick components are detected, we compute the lag between the two
components, $\Delta v=v_{thin}-v_{thick}$, and plot this lag as a
function of radius in Fig.~\ref{summary}.  The 14 fields for which a
thick disc component is reliably detected cover a range of radii from
15.2 to 39.6 kpc. In the top panel of Fig.~\ref{summary}, we can see
that the lag between the two components does not appear to increase
with distance from the centre of M31, and shows an average lag of
$\langle\Delta v\rangle=46.0\kms$.

\begin{figure*}
\begin{center}
\includegraphics[angle=0,width=0.3\hsize]{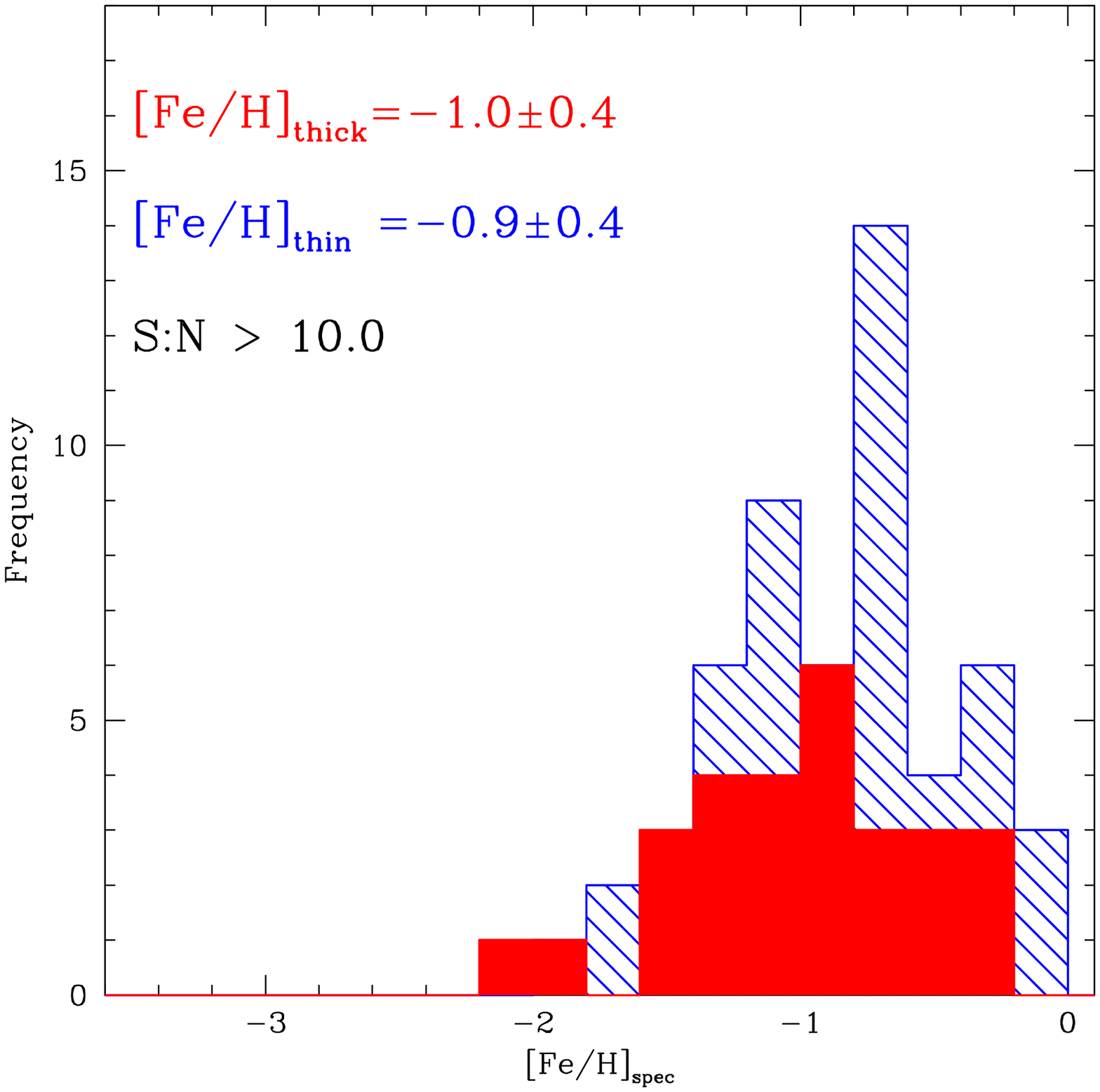}
\includegraphics[angle=0,width=0.3\hsize]{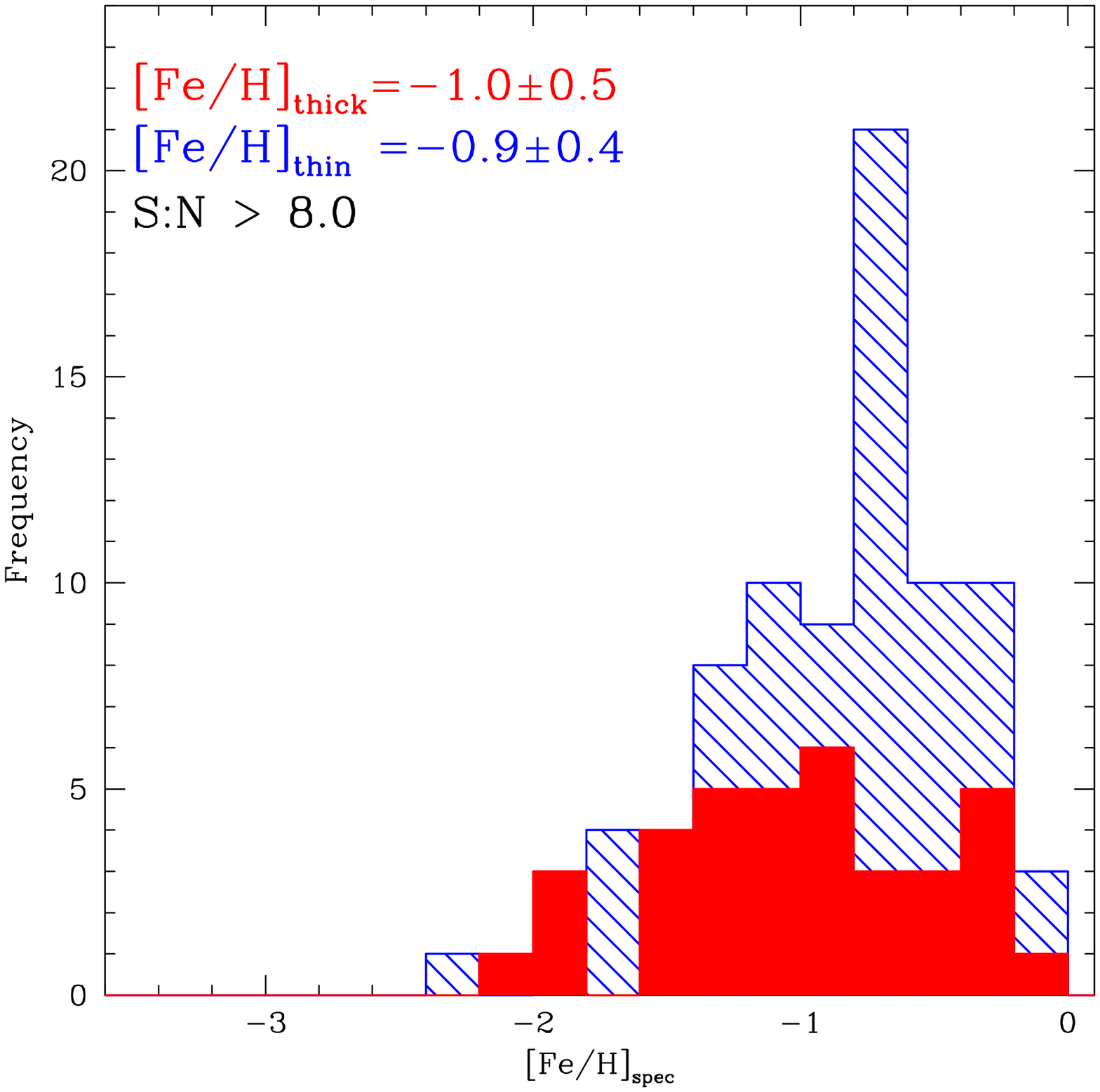}
\includegraphics[angle=0,width=0.3\hsize]{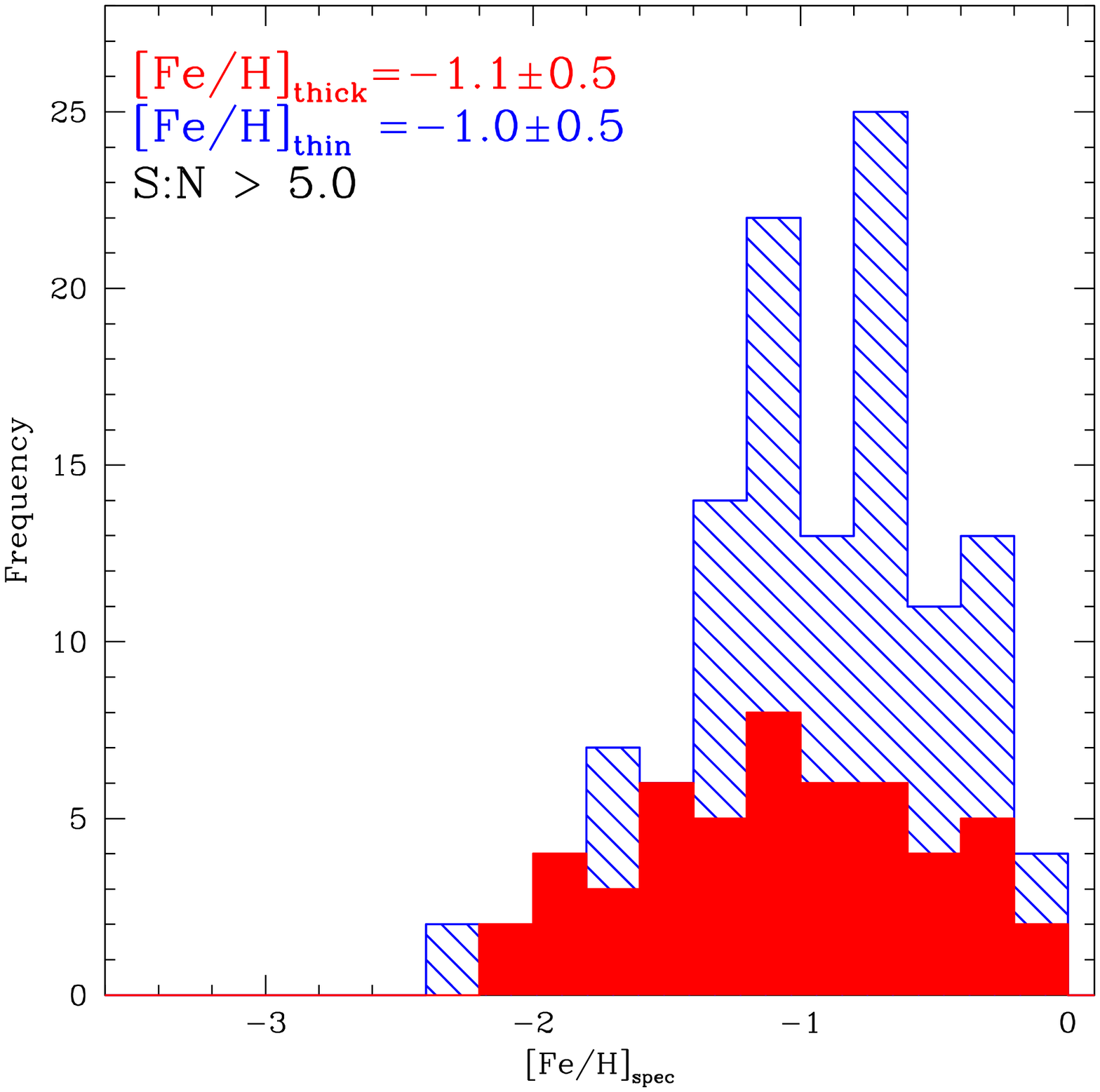}
\caption{Here we display the spectroscopic MDF for all stars in our
  thin and thick components (shown as blue hatched and red solid
  histograms respectively) In the left panel, we show the MDF using
  all stars for which metallicities can be reliably measured (i.e.
  S:N$\ge10$\AA$^{-1}$, and the middle and right panels apply lower
  quality cuts (S:N$\ge8$\AA$^{-1}$ and S:N$\ge5$\AA$^{-1}$). For our
  lower S:N cuts, we note that the median [Fe/H] values for both
  populations remain similar, and the dispersions (inter-quartile
  range) begin to increase, losing some of the detail of the shape of
  the MDF. In all cases the median [Fe/H] of the thick disc is more
  metal poor than the thin by $\sim0.1$ dex.}
\label{spechist}
\end{center}
\end{figure*}

\begin{figure}
\begin{center}
\includegraphics[angle=0,width=0.99\hsize]{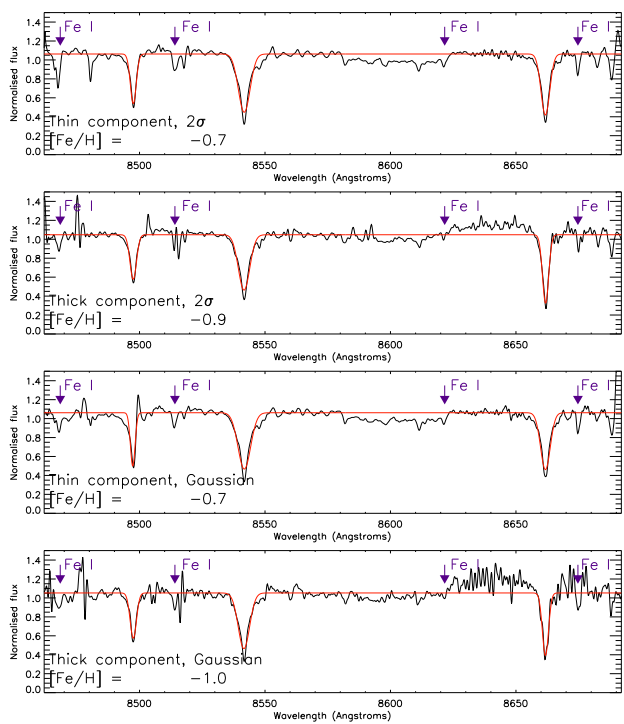}
\caption{Composite spectra of both the thin disc and thick component,
  using both 2$\sigma$ (top panel) and Gaussian (lower panel) cuts to
  isolate the thick component. The composite spectra are constructed
  from stars within the selection regions that possess a S:N $\geq$
  3.0\AA$^{-1}$.  We find that the average metallicity for the thin
  component is more metal rich than the thick by 0.2--0.3 dex. These
  results are consistent with composites formed from spectra with S:N
  $\leq$ 3\AA$^{-1}$ and with our field-by-field metallicity estimates
  (Fig.~\ref{summary}). We also show the locations of a number of Fe I
  lines present in these spectra.}
\label{spectra}
\end{center}
\end{figure}
   
We also plot the dependence of velocity dispersion, $\sigma_{thin}$ and
$\sigma_{thick}$, for both components with radius in the middle panel
of Fig.~\ref{summary}. For the thin disc, we fit both a constant
relation and a single power law to the data. The linear power law
suggests a decrease in dispersion with radius, with a gradient of
-0.87$\kms\kpc^{-1}$, however this fit is not statistically better
than a constant fit, with an average dispersion of
$\sigma_{thin}=31.6\pm1.1\kms$ (reduced $\chi^2$ of 5.6 vs. 5.2). We
do the same for our thick disc results, and we find that a linearly
decreasing profile where
$\sigma_{thick}=-0.8(\pm0.2)R_{proj}+66.1(\pm5.8)$ has a marginally better
fit to the data than a fit with no evolution, however the difference
in negligible (reduced $\chi^2$ of 1.2 vs. 1.4), and deemed
insignificant in a $\chi^2$ significance test. Even if we were to
accept this fit as preferred, we note that the two outermost fields
situated at 34.2 and 39.6~kpc, are perhaps the driving force in the decreasing
dispersion seen in our thick disc component. As these field are the
furthest out in our survey, they also suffers from the greatest chance
of halo contamination in our sample, and therefore could be
unreliable. If we exclude these final points from the fit, we find that
$\sigma_{thick}$ is best fit with no evolution as a
function of radius, with an average dispersion of 50.8$\pm1.9\kms$. We
therefore conclude that our data cannot tell us anything reliable
about the dependence of these kinematic properties with radius, and
allow us to merely calculate the average kinematics of both
components.

\subsubsection{Scale length of the thin and thick disc}

To determine the scale lengths of our two disc components, we need to
calculate the number density of thin and thick disc stars within our
DEIMOS field of view. There are 2 complications we must consider
before we proceed. Firstly, the two components are not completely
distinct from one another, and in all fields, we observe some
overlap. Secondly, owing to our selection criteria (discussed in
\S~3), we prioritise stars of certain colours and magnitudes above
others, and this must be considered when calculating densities on a
field-by-field basis.

We determine the number of stars associated with the extended thin
disc, $n_s$, in each of our fields by integrating the Gaussian we have
fit to this component. To determine the density of stars contained in
the thin disc, we multiply $n_s$ by the total number of available
target stars within our DEIMOS field that fall within our selection
criteria, $n_t$, and divide this by the total number of stars that
were observed with our DEIMOS mask, $n_o$. We then subtract the
density of background stars $n_b$, which is computed from a number of
fields on the edge of our survey region;
i.e. $\rho_*=n_sn_t/n_o-n_b$. To account for our prioritised selection
technique, we perform this calculation separately for our priority A
and priority B stars, then combine these measurements. We repeat this
calculation for the thick disc. We plot the results in
Fig.~\ref{scale}, where we apply a weighted least-squares exponential
fit to our data points, and determine $h_r=7.3\pm1.1$~kpc for the thin
disc and $h_r=8.0\pm1.2$~kpc for the thick. Comparing this to previous
calculations for the scale length of the thin and extended discs, we
find that the extended disc has a larger scale length than the
exponential thin disc, ($5.1\pm0.1$~kpc, I05). The value of 7.3~kpc
that we derive is slightly higher than that derived in I05 of
6.6$\pm0.4$~kpc, and with much larger error bars but the two are
consistent within their $1\sigma$ uncertainties. The difference
between the two values can be attributed to two factors. Firstly, in
I05, they included fields from the NE of the galaxy, plus fields
located away from the major axis, where we have have sampled fields
solely from the SW major axis. Secondly, in I05 they did not fully
address any biases that may have been introduced by our two-tiered
prioritisation system. Finally, we note that the thick disc appears to
be more radially extended than either the thin or extended disc,
although it is consistent with the scale length of the extended disc
within its $1\sigma$-errors.

In previous work \citet{yoachim06} (hereafter YD06) measured
the scale lengths of 34 edge-on disc galaxies using a photometric
fitting technique, and found that the scale lengths of the thick discs
were larger than those of the thin discs by a factor of$\sim1.3$. We
plot their results in the left panel of Fig.~\ref{height}, and overlay
a linear relation with a gradient of 1.3. We add to this our results
for M31, using an average value for the thin disc from the range of
scale lengths derived for the thin and extended discs (5.1--7.3 kpc)
of 6.3~kpc, and our calculated value of 8.0~kpc. We also overplot the
result for the MW (using \citealt{juric08} values of 2.6 and 3.6 kpc
for thin and thick discs respectively). and note that M31 sits in
excellent agreement with this relation.

\begin{table*}
\begin{center}
\caption{Kinematic properties of thin and thick disc components}
\label{kprops}
\begin{tabular}{lccccccc}
\hline
Field & v$_{thin,lag}$ ($\kms$) & $\sigma_{thin} (\kms)$ & v$_{thick, lag}$ ($\kms$) & $\sigma_{thick} (\kms)$ & $\sigma_{conf}$ & [Fe/H]$_{thin}$ & [Fe/H]$_{thick}$\\
\hline
228Dis &  6.8$\pm5.0$  & 55.2$\pm3.2$ & -141.9$\pm7.5$ & 41.2$\pm11.8$ & 37.0 & -0.8$\pm$0.1 & -1.0$\pm$0.1\\\
227Dis & -3.7$\pm10.8$ & 68.7$\pm3.3$ & N/A            & N/A           & 2.8  & -0.7$\pm0.2$ &N/A\\
166Dis & -7.5$\pm6.9$  & 48.9$\pm4.6$ & N/A            & N/A           & 0.5 & -0.8$\pm$0.2 & N/A\\
106Dis & 11.6$\pm$8.2  & 47.4$\pm5.2$ & -85.0$\pm8.2$  & 52.5$\pm4.1$  & 3.5 & -0.9$\pm$0.3 &  -1.0$\pm0.3$\\
105Dis & -16.4$\pm7.2$ & 32.0$\pm4.1$ & -54.8$\pm8.7$  & 51.0$\pm5.2$  & 4.0 & -0.7$\pm$0.2 & -1.0$\pm$0.2\\
224Dis & -34.4$\pm6.2$ & 24.3$\pm2.5$ & -63.8$\pm10.2$ & 55.0$\pm12.5$ & 5.1 & -0.9$\pm$0.1 & -1.1$\pm$0.2\\
232DiS & -37.2$\pm3.0$ & 35.0$\pm3.8$ & N/A            & N/A           & 2.0 & -0.8$\pm$0.1 & N/A\\
104Dis & -37.9$\pm6.5$ & 30.0$\pm4.4$ & -114.3$\pm9.5$ & 53.3$\pm7.2$  & 9.0 & -0.7$\pm$0.2 & -0.9$\pm$0.1\\
220Dis & -12.5$\pm5.0$ & 39.6$\pm2.9$ & -64.8$\pm12.1$ & 54.7$\pm10.9$ & 26.7 & -1.1$\pm$0.3 & -1.6$\pm$0.3\\
213Dis & -10.9$\pm3.2$ & 31.9$\pm6.3$ & -80.5$\pm10.0$ & 55.0$\pm8.8$  & 39.2 & -0.8$\pm$0.3 & -1.5$\pm$0.3\\
102Dis & -18.8$\pm7.1$ & 22.2$\pm5.2$ & -61.5$\pm11.6$ & 48.1$\pm5.6$  & 47.1 & -0.9$\pm$0.2 & -1.0$\pm$0.2\\
231Dis &   4.7$\pm15.2$& 22.9$\pm4.2$ & -35.9$\pm7.8$  & 53.4$\pm7.2$  & 43.4 & -0.9$\pm$0.1 & -1.1$\pm$0.2\\
223Dis &  -8.1$\pm5.9$ & 20.1$\pm4.9$ & -55.7$\pm9.9$  & 51.1$\pm6.8$  & 14.7 & -0.7$\pm$0.2 & -1.0$\pm$0.3\\
101Dis & -16.2$\pm3.9$ & 35.8$\pm5.8$ & -57.3$\pm8.6$  & 44.9$\pm6.2$  & 3.1 & -1.1$\pm$0.2 & -1.3$\pm$0.3\\
222Dis & -25.9$\pm$4.8 & 33.6$\pm2.9$ & N/A            & N/A           & 0.8 & -0.7$\pm$0.1 & N/A\\
221Dis & -27.5$\pm5.5$ & 35.0$\pm3.6$ & -121.2$\pm7.4$ & 52.2$\pm6.3$  & 27.1 & -1.1$\pm$0.2 & -1.4$\pm$0.2\\
50Disk & -16.9$\pm8.1$ & 34.2$\pm5.9$ & -149.2$\pm13.2$& 42.3$\pm7.2$  & 12.6 & -1.0$\pm$0.2 & -1.2$\pm$0.2\\
107Ext & -55.2$\pm6.9$ & 24.8$\pm4.4$ & N/A            & N/A           & 2.4 & -1.0$\pm$0.3 & N/A\\
w11old & -55.1$\pm10.2$& 28.8$\pm5.6$ & N/A            & N/A           & 1.4 & -1.0$\pm$0.3 & N/A\\
167Hal & -72.6$\pm8.3$ & 22.9$\pm5.4$ & -120.3$\pm12.2$& 35.1$\pm7.8$  & 23.5 & -0.9$\pm$0.3 & -1.0$\pm0.3$\\
148Ext & -17.2$\pm7.0$ & 41.3$\pm6.6$ & -154.5$\pm11.0$& 25.7$\pm5.1$  & 28.2 & -0.9$\pm$0.3 & -1.0$\pm$0.2\\
\hline
\end{tabular}
\end{center}
\end{table*}

\begin{table}
\begin{center}
\caption{Average properties of thin and thick disc components derived in this work}
\label{avprops}
\begin{tabular}{lcccc}
\hline
Component & $\sigma_v (\kms)$ & $h_r$ (kpc) & z$_0$ (kpc)& [Fe/H]$_{spec}$\\
\hline
Thin disc  & 35.7$\pm1.0$ & 7.3$\pm1.1$ & 1.1$\pm0.2$ & -0.7$\pm0.05$ \\
Thick disc & 50.8$\pm1.9$ & 8.0$\pm1.2$ & 2.8$\pm0.6$ & -1.0$\pm0.1$ \\
\hline
\end{tabular}
\end{center}
\end{table}

\subsubsection{Inferring the scale heights of the thin and thick discs}   

Owing to the inclination of M31, we are unable to measure the height
of either the thin or thick disc components directly. No photometric
excess above a typical bulge or extended disc profile is observed when
performing minor axis star counts \citep{irwin05}, suggesting that
these components dominates the surface profile out to large radii. In
order to infer probable scale heights for both components, we make use
of the properties of the 34 edge-on galaxies measured by YD06. As the
scale lengths and heights of both thin and thick discs in each of
these galaxies were derived, it is possible for us to search for a
relation between the scale length, $h_r$, and scale height, $z_0$ of
each component. In the central panel of Fig.~\ref{height}, we plot
$h_r$ vs. $z_0$ for the YD06 sample as well as for the MW
\citep{ivezic08}, and fit it with a linear relation, on which we force
an intercept of (0,0). We find that the data are well fit with a
gradient for this relation of 0.18$\pm0.04$, though there is
significant scatter beyond $\sim9$~kpc. From this, we deduce
$z_0$=1.1$\pm0.2$~kpc for the M31 extended disc (using $h_r$=7.3
kpc). We repeat this for the thick disc (shown in the right panel of
Fig.~\ref{height}) and find that these values are well fit with a
linear relation of gradient 0.35$\pm0.06$, giving us
$z_0$=2.8$\pm0.6$~kpc for the M31 thick disc. If these values are
correct, then not only are the discs of M31 more radially extended
than those of the MW by a factor of $\sim2-3$, they are also
significantly thicker.

\subsubsection{Contrast of the thin and thick discs}

In the previous sections, we have derived the density in each field of
both our components as a means to determine the scale lengths. We now use
these densities to work out how much of the total (disc related)
stellar population is contained within either component. There are
several caveats to such a comparison that should be
mentioned. Firstly, our sampling of the field is likely to have
an effect on our field-to-field estimates of the stellar density
(which we discuss further in \S~5). Secondly, as the disc is not
observed edge-on, we are measuring a 2D projection of the
densities which is difficult to interpret. We also note that the
measurement errors associated with the densities of each field (shown
in table~\ref{densities}) are significant (of order $\sim50$\%). 

With this in mind, we find that, on average, the thick disc component
accounts for $35$\% of the total stellar density, with an
inter-quartile range of $\pm$10\%. In the Milky Way, we know that the
thick disc contributes to $\sim10$\% of the stellar density in the
solar neighbourhood, and accounts for $\sim1/3$ of the {\it total}
disc mass \citep{juric08,schonrich09a}, comparable to what we derive here. 

From our calculated contrasts and individual density profiles for the
thin and thick discs, we can estimate the mass contained within the
thick disc component using values for the mass of the thin disc from
the literature. From our analysis above, we have determined that the
thick disc contributes 35$\pm10$\% of the {\it total} stellar density,
meaning the thick:thin disc density ratio is of order 55$\pm15$\%. We
can also estimate this fraction by integrating our stellar density
profiles (Fig.~\ref{scale}) over the limits of our data, and from this
we calculate a thick:thin disc density ratio of $\sim$65\%, which is
in good agreement with our contrast estimate. If we assume that both
discs are composed of similar stellar populations, we can set the mass
ratio between the disc to be equivalent to the density ratio. In
\citet{yin09}, they quote a total mass for the thin stellar disc of
$M_{*,thin}=5.9\times10^{10}\msun$, calculated from the mass models of
\citet{widrow03} and \citet{geehan06}. From this we estimate that the
total mass of the M31 thick disc lies in the range
2.4$\times10^{10}\msun<M_{*,thick}<4.1\times10^{10}\msun$. As we are
unable to determine the full radial and underlying luminosity profile
for the thick disc, these values are obviously prone to large errors
introduced by our simplifying assumptions, and the mass of the thick
disc may be lower than our quoted range. For example, if we just
compare the number of stars we detect in the thin disc throughout our
entire sample with the number we detect in the thick disc by
integrating the fitted Gaussians in Fig.~\ref{fields} we find a
thick:thin disc ratio of 20\%. If we assume this value for the ratio
of the masses between the components, our lower limit on the thick
disc is reduced to $M_{*,thin}=1.2\times10^{10}\msun$. A future study
of the thick disc which includes fields from the entirety of our
survey will help us to better constrain both the radial profile and
mass of this component.

\begin{table}
\begin{center}
\caption{Field by field densities of thin and thick disc stars}
\label{densities}
\begin{tabular}{lcccccc}
\hline
Field & $\rho_{* thin}$ (*/arcmin) & $\rho_{* thick}$ (*/arcmin) \\
\hline
228Dis & 57.6$\pm$22.3 & N/A\\
227Dis & 56.9$\pm$22.3 & N/A\\
166Dis & 20.4$\pm$9.4  & N/A\\
106Dis & 36.2$\pm$15.3 & 23.6$\pm$10.5\\
105Dis & 45.3$\pm$18.6 & 41.0$\pm$16.3\\
224Dis & 20.1$\pm$8.4  & 17.1 $\pm$7.1\\
232DiS & 35.9$\pm$15.7 & N/A \\
104Dis & 35.7$\pm$14.6 & 6.6$\pm$4.5\\
220Dis & 19.9$\pm$8.9  & 13.5$\pm$6.4\\
213Dis & 23.0$\pm$12.2 & 9.8$\pm$3.0\\
102Dis & 24.0$\pm$12.8 & 8.9$\pm$4.5\\
231Dis & 13.0$\pm$6.6  & 16.6$\pm$8.1\\
223Dis & 15.1$\pm$6.7  & 12.2$\pm$5.6\\
101Dis & 26.5$\pm$12.9 & 11.1$\pm$4.8\\
222Dis & 4.9$\pm$ 2.1  & N/A \\
221Dis & 14.4$\pm$6.4  & 4.8$\pm$3.3\\
50Disk & 8.1$\pm$3.9   & 2.0$\pm$0.6\\
107Ext & 9.6$\pm$7.8   & N/A\\
w11old & 9.0$\pm$6.5   & N/A\\
167Hal & 1.3$\pm$0.8   & 0.8$\pm0.5$\\
148Ext & 1.1$\pm$0.9   & 0.7$\pm$0.5\\
\hline
\end{tabular}
\end{center}
\end{table}

\subsection{Spectroscopic metallicities}

In this section we present the spectroscopic values of [Fe/H], both
for individual stars, and for the composite spectrum of each
component. Measuring individual metallicities from the Ca II triplet
for the stars in our survey, with S:N typically between
5--15\AA$^{-1}$, is quite problematic. In \citet{battaglia08}, they
show that the `best case' errors in measuring the equivalent widths
of the Ca II lines, $\Delta EW$, scale with S:N as:

\begin{equation}
\Delta EW=\frac{\sqrt{1.5\times FWHM}}{S:N}
\end{equation}

\noindent assuming no contamination from residual sky lines and no covariance noise, where
FWHM is the full width at half-maximum of the CaT lines which is
typically 2--3\AA. Using this equation, we can determine the average
errors in [Fe/H] for stars in our sample at different S:N, and we find
that for spectra with S:N of (5, 8, 10)\AA$^{-1}$, the errors in their
calculated metallicity are of order (0.6, 0.4, 0.3) dex. To
demonstrate the effects of these large errors on the metallicity
distribution function (MDF) of our sample, we present histograms of
the individual spectroscopic metallicities for three quality cuts, one
at S:N$\ge10$\AA$^{-1}$ (left panel), one with S:N$\ge8$\AA$^{-1}$
(middle panel) and one with S:N$\ge5$\AA$^{-1}$ (right panel) in
Fig.~\ref{spechist}. In both plots, the blue hatched histogram
represents our thin disc sample and the filled red histogram
represents our thick disc population. For our higher quality spectra,
we calculate a median metallicity for the thin disc of [Fe/H]=-0.9
with a dispersion of 0.4 dex (from the inter quartile range, IQR). For
our thick disc population we calculate a median of [Fe/H]=-1.0 also
with a dispersion of 0.4 dex. We note that the distributions of both
populations deviate from a Gaussian distribution, with a kurtosis of
-0.6 and -0.3 for thin and thick discs respectively, implying a broad
peaked distribution, with narrow tails. Both distributions are skewed
towards lower metallicity with skewness $\alpha=-0.4$ and
$\alpha=-0.3$ for thin and thick disc. For our lowest S:N cut,
however, much of this information is lost. While the median [Fe/H]
remains very similar with [Fe/H]=-1.0 for the thin disc and
[Fe/H]=-1.1 for the thick, the distributions begin to broaden, with
dispersions of 0.5 dex for both populations, and present almost no
skew ($\alpha=-0.1$ and $\alpha=-0.2$ for thin and thick disc). This
shows that by including data with larger measurement errors, we wash
out our MDF considerably, and lose any meaningful information. As a
sanity check, we compare the MDF for all stars with
S:N$\ge10$\AA$^{-1}$ with one for stars with S:N$\ge15$\AA$^{-1}$, and
find both the median values of [Fe/H] and general distributions to be
comparable. We note that by requiring such a high S:N cut on our
individual measurements of [Fe/H], we bias our sample towards more
metal-rich stars as these will have intrinsically stronger Ca II
lines.

Owing to the large errors associated with these measurements, this
analysis provides a crude indication of the metallicities of both
discs, and so to get a more accurate estimate of the average
metallicities of both populations we construct composite spectra for
both components (using both the 2$\sigma$ and Gaussian velocity cuts)
by co-adding the individual spectra of all stars with
S:N$>3$\AA$^{-1}$, weighted by their S:N values. The resulting S:N of
the composite is much greater than the individual spectra
(S:N$\sim60-100$ cf. S:N$\sim3-25$), allowing a better fit to the CaII
lines. We use a cut of S:N$>3$\AA$^{-1}$ as below this the velocity
uncertainties of our stars begin to significantly increase (as
discussed in \S~3). As we shift all spectra to the rest frame before
co-adding, including spectra where the velocity is uncertain could
smear out the Ca II lines, resulting in an over-estimate of [Fe/H] for
the composite. We note that the results from our composites are only
indicative of an average metallicity for each component, and can tell
us nothing about the metallicity dispersion for the discs. We display
the resulting composites in Fig.~\ref{spectra}. The top 2 panels show
the thin and thick spectra for the 2$\sigma$ velocity cuts, while the
bottom 2 panels show the same, but for the Gaussian velocity cuts. In
the case of the 2$\sigma$ cuts, our thin composite comprises 511 stars
that match our kinematic and quality criteria, while our thick
composite is constructed from 78 stars. For our Gaussian cuts, these
numbers fall to 380 and 52 stars respectively. We find an offset of
order 0.2 dex between the thick and thin components for the 2$\sigma$
cuts, with the thick disc being more metal poor at [Fe/H]$=-0.9\pm0.1$
compared with [Fe/H]$=-0.7\pm0.05$ for the thin, inconsistent within
their respective 1$\sigma$ errors. For our Gaussian cuts, we find the
thick disc to be more metal poor, giving us a larger difference in
metallicity between the two components of 0.3 dex (with
[Fe/H]$=-1.0\pm0.1$ for the thick disc compared with
[Fe/H]$=-0.7\pm0.05$ for the thin), although the two results for the
thick disc are consistent within their 1$\sigma$ errors. We also note
that we are liable to experience non-negligible thin disc
contamination of our thick disc component, which could cause us to
over estimate the average [Fe/H], so the true difference could be
larger still. We note that these results are consistent with
performing the same analysis on composites constructed from spectra
with S:N$>10$\AA$^{-1}$.

Finally, the continuum fit to the third line in our composite spectra,
particularly for our thick disc selection, gives us some cause for
concern. Could this metallicity difference we derive be driven by poor
continuum fitting in this region of the spectrum? To investigate this,
we analyse the [Fe/H] for the thin and thick discs again, using solely
the first two lines (CaII$_{8498}$ and CaII$_{8542}$. In the case of
our simple $2\sigma$ cut, this narrows our difference in metallicity
slightly from 0.2 dex to 0.15 dex, with [Fe/H]$=-0.85\pm0.1$ compared
with [Fe/H]$=-0.7\pm0.05$ for the thick and thin discs
respectively. However, in the case of our Gaussian cuts, which are
arguably less affected by cross contamination between the components,
the metallicity difference of 0.3 dex persists.

We also perform this composite analysis on a field-by-field basis. The
results of this analysis, shown in Table~\ref{kprops} are again, less
accurate than our overall composite, but they suggest a similar offset in
metallicity exists in the thin and thick components in each field. We
plot this result as a function of radius in the lower panel of
Fig.~\ref{summary}. We find no evidence for any evolution of
metallicity with radius.

\begin{figure*}
\begin{center}
\includegraphics[angle=0,width=0.99\hsize]{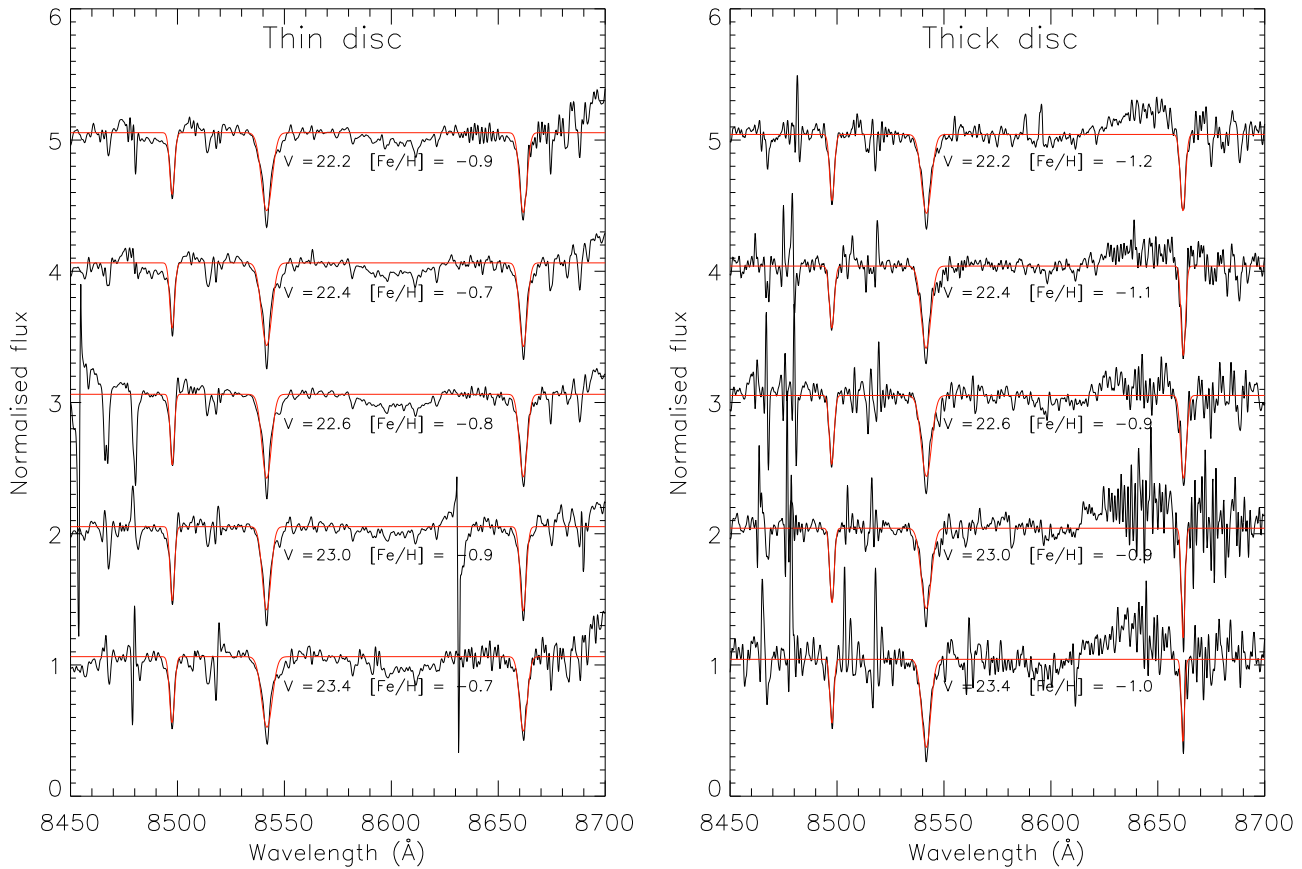}
\caption{Composite spectra of our thin (left panel) and thick (right panel) disc samples, binned in $V$-band magnitude. Each bin spans 0.2 mags. We see that in each case, the thick disc is more metal-poor than the thin disc by $\sim0.2$ dex. The errors in the values of [Fe/H] for these composites ranges from 0.1--0.3 dex.}
\label{binned}
\end{center}
\end{figure*}

A slight concern in ascertaining the metallicity of a population from
a composite spectrum arises from inaccuracies in the estimate that
come from combining spectra with different effective
temperatures and $V$-band magnitudes, as the derived metallicities are
weakly dependent on the apparent $V$-band colours of the stars. The
rms dispersion in the $V$-band magnitudes within our sample are small
($<0.5$ mag for both thin and thick discs) as we are sampling only a
small region of the tip of the RGB, so the error introduced by this
effect will be very small. However, to further assess this, we
separate our thin and thick disc spectra into bins of 0.2 mags in the
$V$-band and create composite spectra for each bin, measuring the
metallicity of each. We show a sample of these spectra in
Fig.~\ref{binned}, labelled with the metallicity and average $V$-band
magnitude. The typical errors in metallicities determined for these
composites ranges from 0.1--0.3 dex. What we see is that the composite
thick disc spectrum in each bin is more metal-poor than the
corresponding thin disc composite. We also find that the average
metallicities for both thin and thick discs agree with those that we
derived from the composites for the entire sample.

\subsection{Photometric Metallicities}

\begin{figure*}
\begin{center}
\includegraphics[angle=0,width=0.3\hsize]{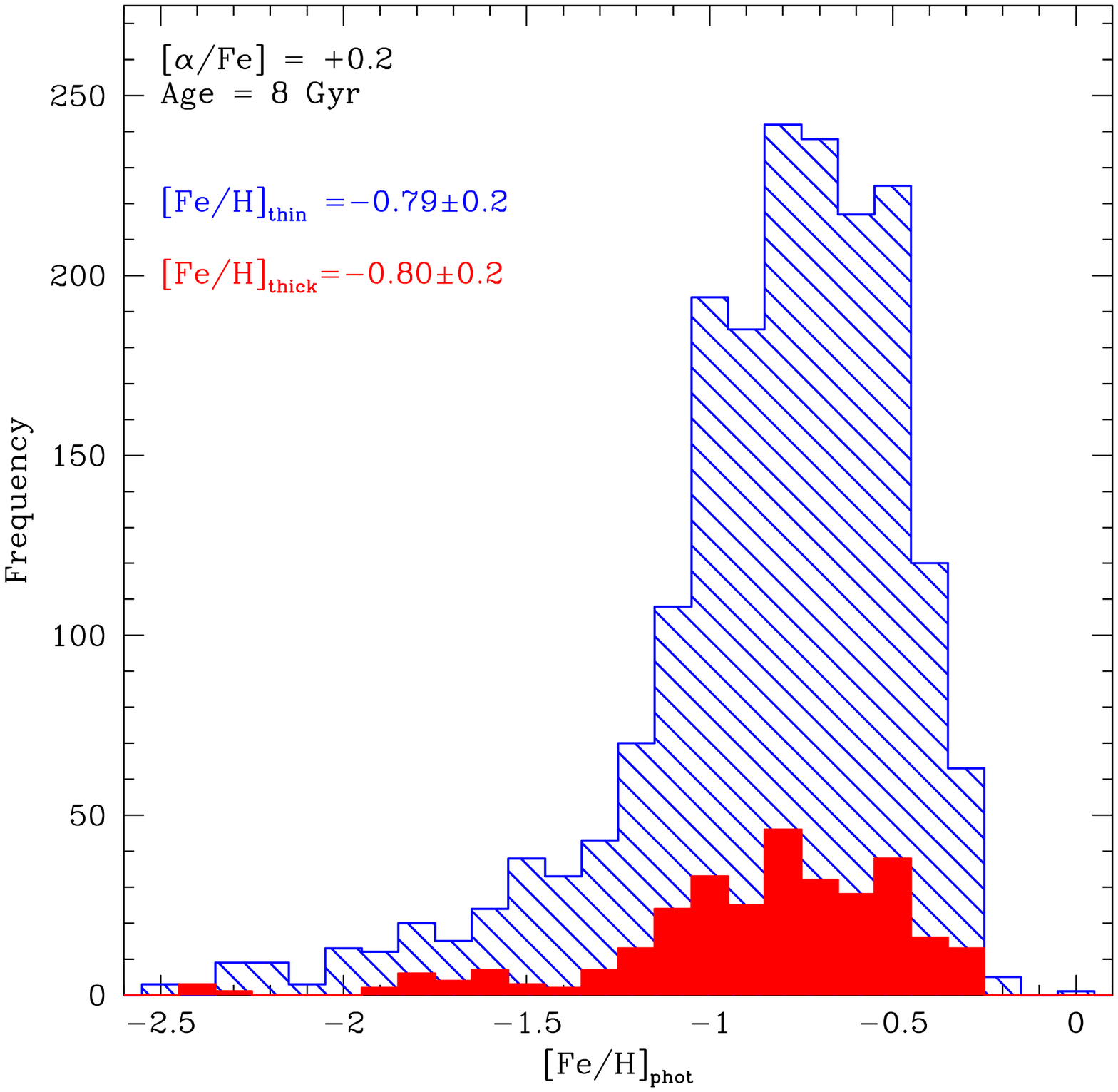}
\includegraphics[angle=0,width=0.3\hsize]{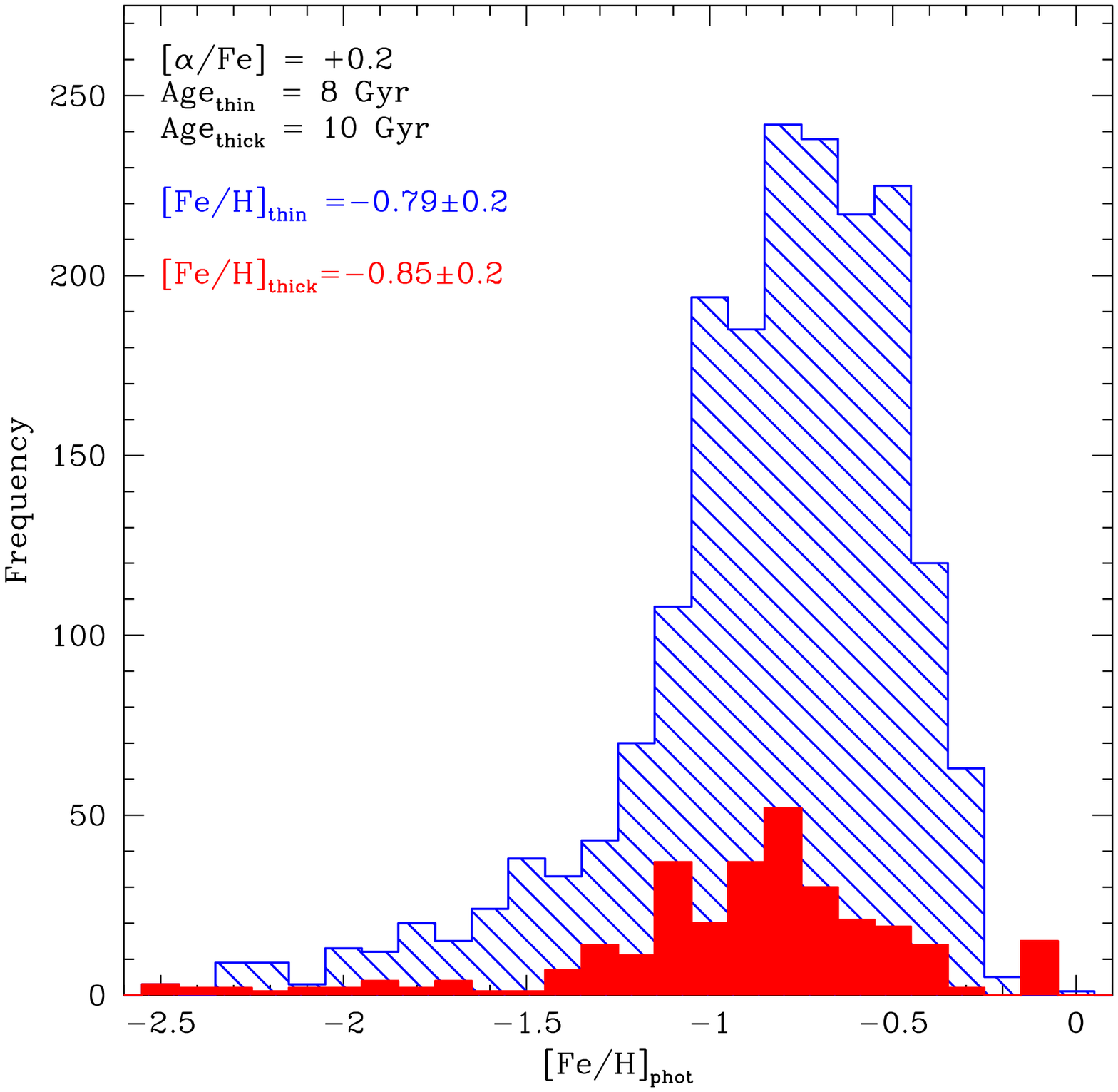}
\includegraphics[angle=0,width=0.3\hsize]{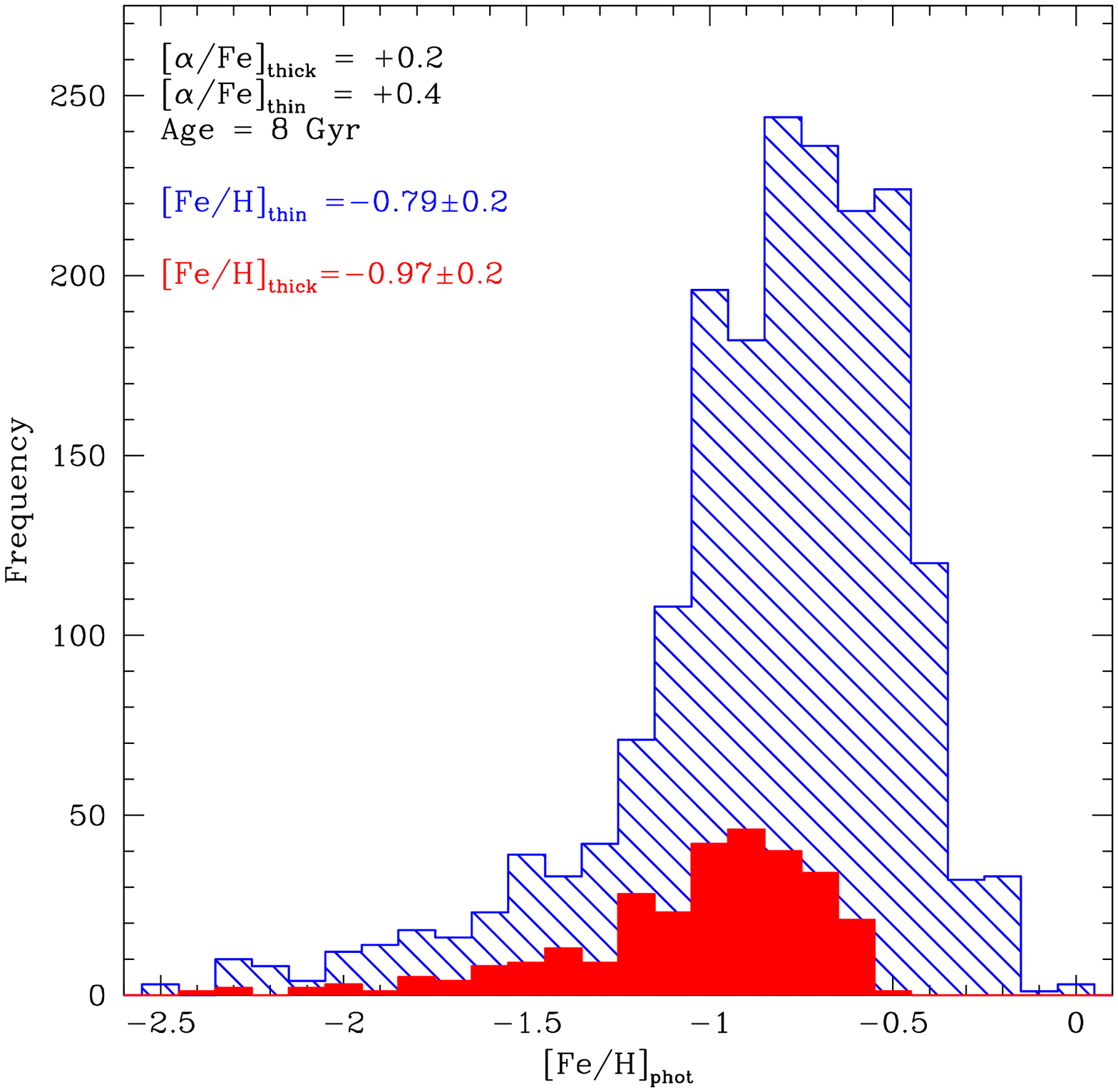}
\caption{Photometric MDFs derived from Dartmouth isochrones of varying
  age and [$\alpha$/Fe] \citep{dart08} for the thin and thick
  components (shown as blue hatched and red filled histograms
  respectively) as defined by our 2$\sigma$ cuts. {\bf Left panel: }
  Analysis of [Fe/H] for thin and disc using [$\alpha$/Fe]=+0.2 and an
  age of 8 Gyrs. We detect no significant differences between the two
  populations, calculating median [Fe/H] and
  [Fe/H]$_{thick}=-0.8\pm0.2$. {\bf Centre panel: } Increasing the age
  of isochrones used to calculate metallicity for the thick disc from
  8 Gyrs to 10 Gyrs. An offset of $\sim0.05$ dex in the average [Fe/H]
  of the two components is observed, with median
  [Fe/H]$_{thick}=-0.85$. The dispersion remains the same as
  before. {\bf Right panel: } Increasing [$\alpha$/Fe] for the thick
  disc from +0.2 to +0.4. An offset of $\sim0.1$ dex between the
  median metallicities of the populations is now observed, with
  [Fe/H]$_{thick}=-0.93$.}
\label{mdf}
\end{center}
\end{figure*}

We inspected the photometric metallicities of our sample using the
Dartmouth isochrone models \citep{dart08}.  We select an age of 8 Gyrs
as the work of \citet{brown06} suggests that the age of the disc in
these outer regions varies between 4 and 8 Gyrs. We use an
$\alpha$-abundance of [$\alpha$/Fe]=+0.2 as it has been shown in
various works (e.g. \citealt{reddy06,abrito10}) that the
$\alpha$-enhancement of thin or extended stellar disc populations
typically ranges between [$\alpha$/Fe]=+0.0 and [$\alpha$/Fe]=+0.2. We
then interpolate between these isochrone models for every star within
our sample to determine its metallicity.  We can then compare the
MDFs for our thin and thick disc sample, selected by both the
2$\sigma$ and Gaussian cuts discussed above. The results of this are
shown in the left panel of Fig.~\ref{mdf}. This figure shows us that
when using this set of isochrones, the MDFs of both populations trace
each other remarkably well. We calculate a median metallicity for each
component and find [Fe/H]$_{thin}=-0.79$ and [Fe/H]$_{thick}=-0.80$,
both with IQRs of 0.2 dex. Neither population has a Gaussian
distribution, with positive kurtosis of +2.2 for both MDFs (i.e. more
peaked, with broader tails), and both populations are skewed towards
lower [Fe/H] with $\alpha\sim-1.2$ for both discs. From this analysis,
one might conclude that the two discs are chemically
indistinguishable. This is in contrast to our findings from the
combined spectra in \S4.2 where we find an offset in the average
metallicities of thin and thick components of 0.2 dex. As our
photometric data are not deep enough to detect the MSTO of these
fields, we are exposed to the age-metallicity-[$\alpha$/Fe] degeneracy
problem. If we analyse our data with isochrones of different ages and
abundances, we find that the individual metallicities we measure
change. Increasing the age by 2 Gyrs has the effect of decreasing
[Fe/H] of a star by $\sim0.05$ dex on average (shown in the centre
panel of Fig~\ref{mdf}) and increasing the abundance from
[$\alpha$/Fe]=+0.2 to [$\alpha$/Fe]=+0.4 reduces [Fe/H] by
$\sim0.1$~dex, (right hand panel, Fig.~\ref{mdf}). The dispersions,
kurtosis and skew remain largely unchanged by these variations. These
findings demonstrate that it may be difficult to discern slight
differences in metallicity (such as the 0.2 dex measured in \S4.2)
using photometric isochrones without knowing the ages and/or
$\alpha$-abundances of the thick and thin disc. Studies of the thin
and thick discs in the MW have shown that the thick disc is both older
and more $\alpha$-enriched than the thin disc
\citep{reddy06,abrito10}, and many of the formation scenarios of thick
discs suggest this could be true for thick discs in general, including
M31. Such differences would certainly affect our derived values of
[Fe/H] for both discs.

\section{Discussion}

In this section we discuss our findings, and comment on the
morphology of this thick component. First we compare our findings with
an expected thin+thick disc population inclined to us along the line
of sight by 77$\deg$, by creating a model of a galaxy with a
thin/extended disc with similar properties to those of M31 that has an
additional thick disc component and analysing it in the same way as
our data. We then compare the M31 thick disc to the MW and the
\citet{yoachim06} sample of thick discs. Finally we comment on the
possible formation mechanisms for this component.

\subsection{Comparison with thin + thick disc model}

To lend confidence to our defining the lagging component we isolate in the above
analysis as a thick disc, we create a simple kinematic model
of a galaxy with properties similar to those of a MW-type galaxy,
which has both a thin and thick stellar disc, and analyse this in the
same manner as our data. This is done as follows; first, we create a
thin stellar disc of 9$\times10^{6}$ stars, randomly generating radii
for each assuming the stars are distributed in an exponential disc
with a scale length equal to that of M31's (6.6 kpc,
I05). We assign each particle with a velocity randomly
drawn from a Gaussian population centred on 0$\kms$ with a velocity
dispersion of 25$\kms$ in the disc frame. We repeat this for our thick
disc component, assuming a thin:thick disc density ratio in M31 that
is equal to that measured in the solar neighbourhood of 9:1
\citep{just10}, giving 1$\times10^6$ stars, and we use a thick disc
scale length of 8.0 kpc, as determined above. For the velocities, we
assume the thick disc lags behind the thin by 50$\kms$ and has the
same velocity dispersion as the MW thick disc, $\sigma=40\kms$
\citep{ivezic08}. We also generate vertical heights within the discs
for both thin and thick populations, assuming MW scale heights for the
discs (300 pc and 1000 pc for thin and thick disc,
\citealt{ivezic08}), $I$-band magnitudes between $25.0\ge I\ge20.3$
(0.1 mags brighter than the tip of the RGB in M31) and angular
positions within the disc. We then convert our disc frame velocities
into heliocentric velocities using the HI rotation curve of
\citet{chemin09}. Finally, we interpret this model in the same way as
our data, by rotating it into the coordinate system of M31 as observed
from the MW (with inclination and PA as discussed in \S~3), and
subtracting off the assumed disc velocity at that position by
interpolating each of our model stars into our average disc velocity
map (Fig.~\ref{velmap}).

\begin{figure*}
\begin{center}
\includegraphics[angle=0,width=0.4\hsize]{velcuts_new.eps}
\includegraphics[angle=0,width=0.4\hsize]{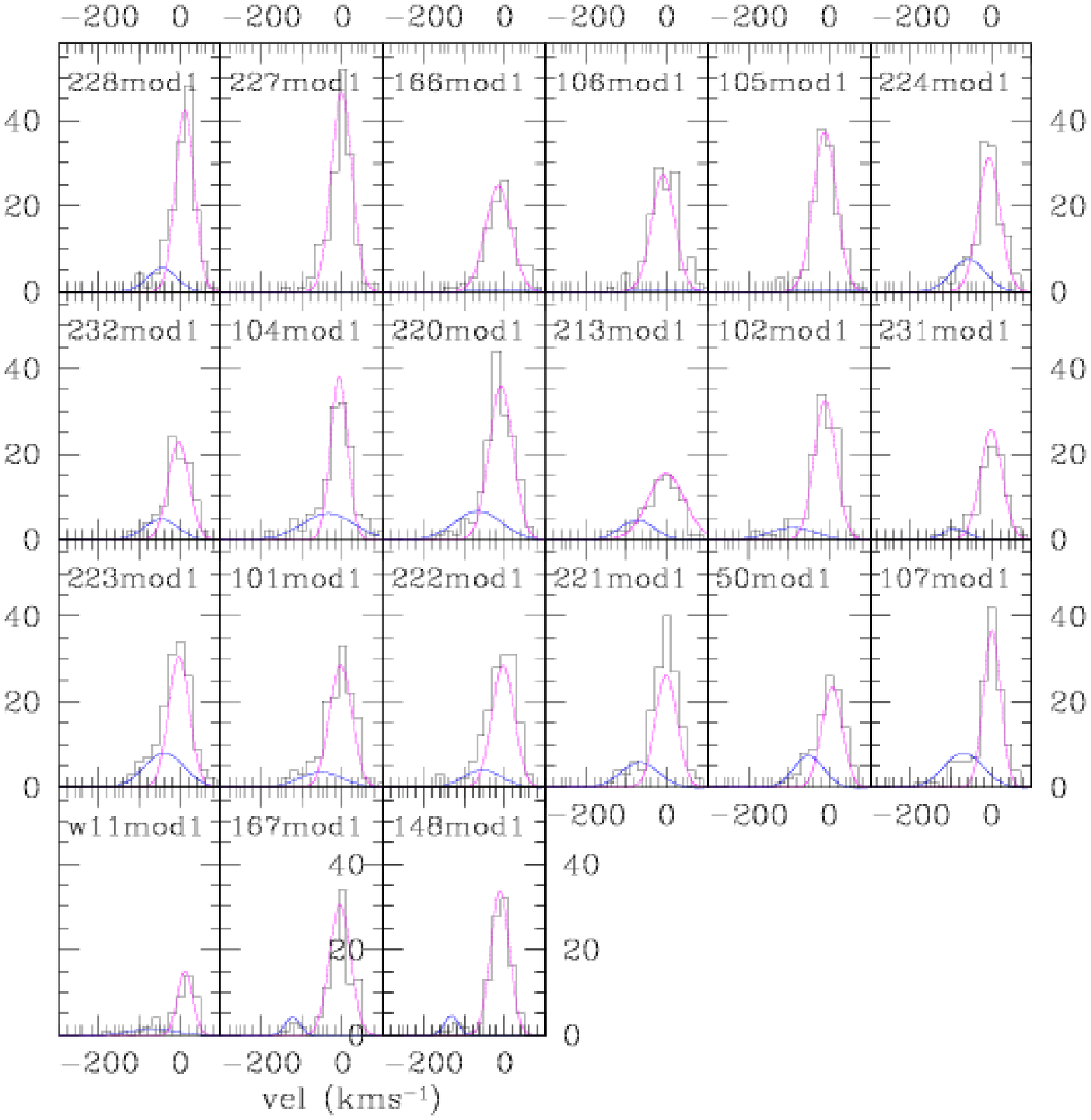}
\includegraphics[angle=0,width=0.4\hsize]{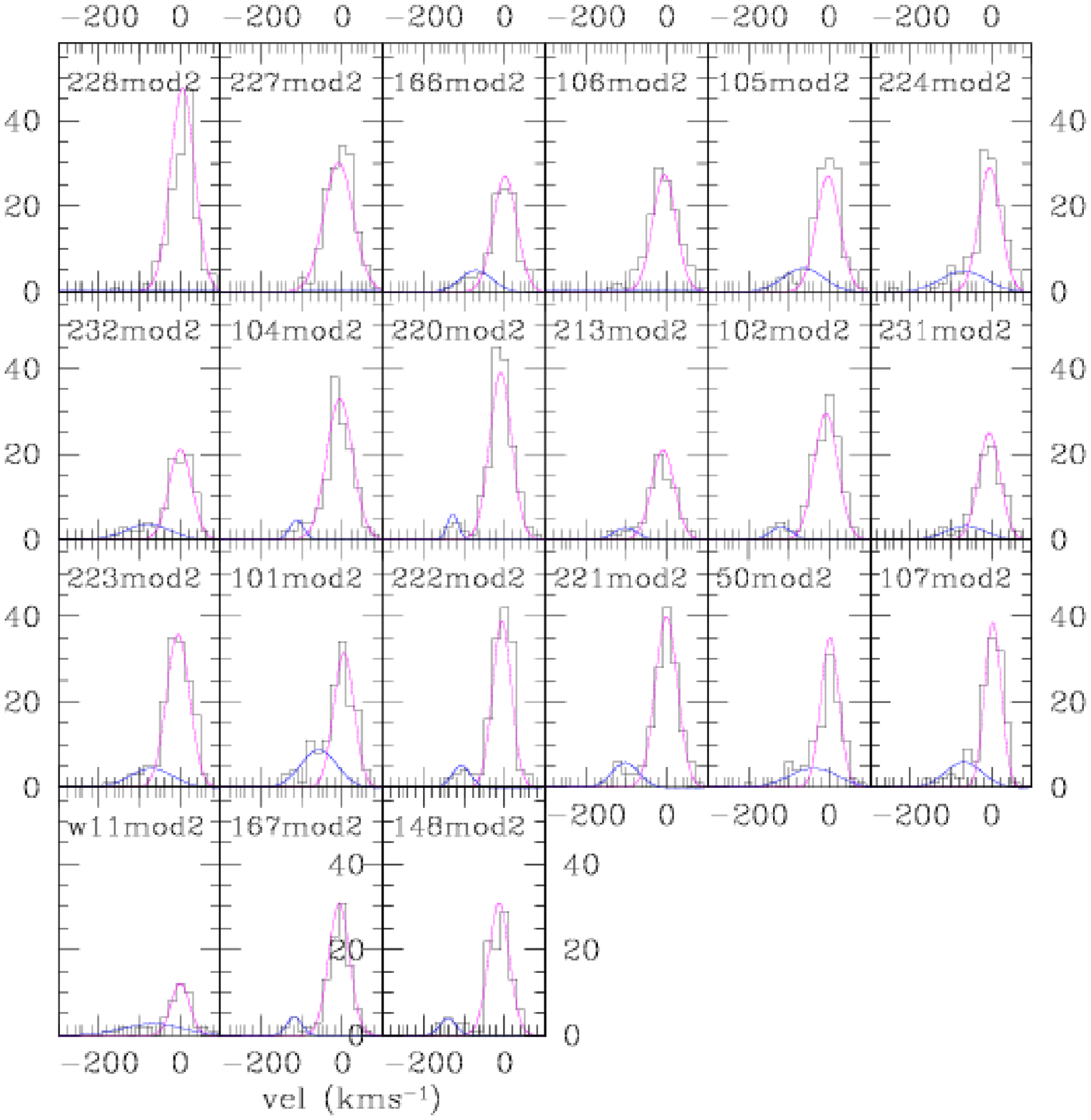}
\includegraphics[angle=0,width=0.4\hsize]{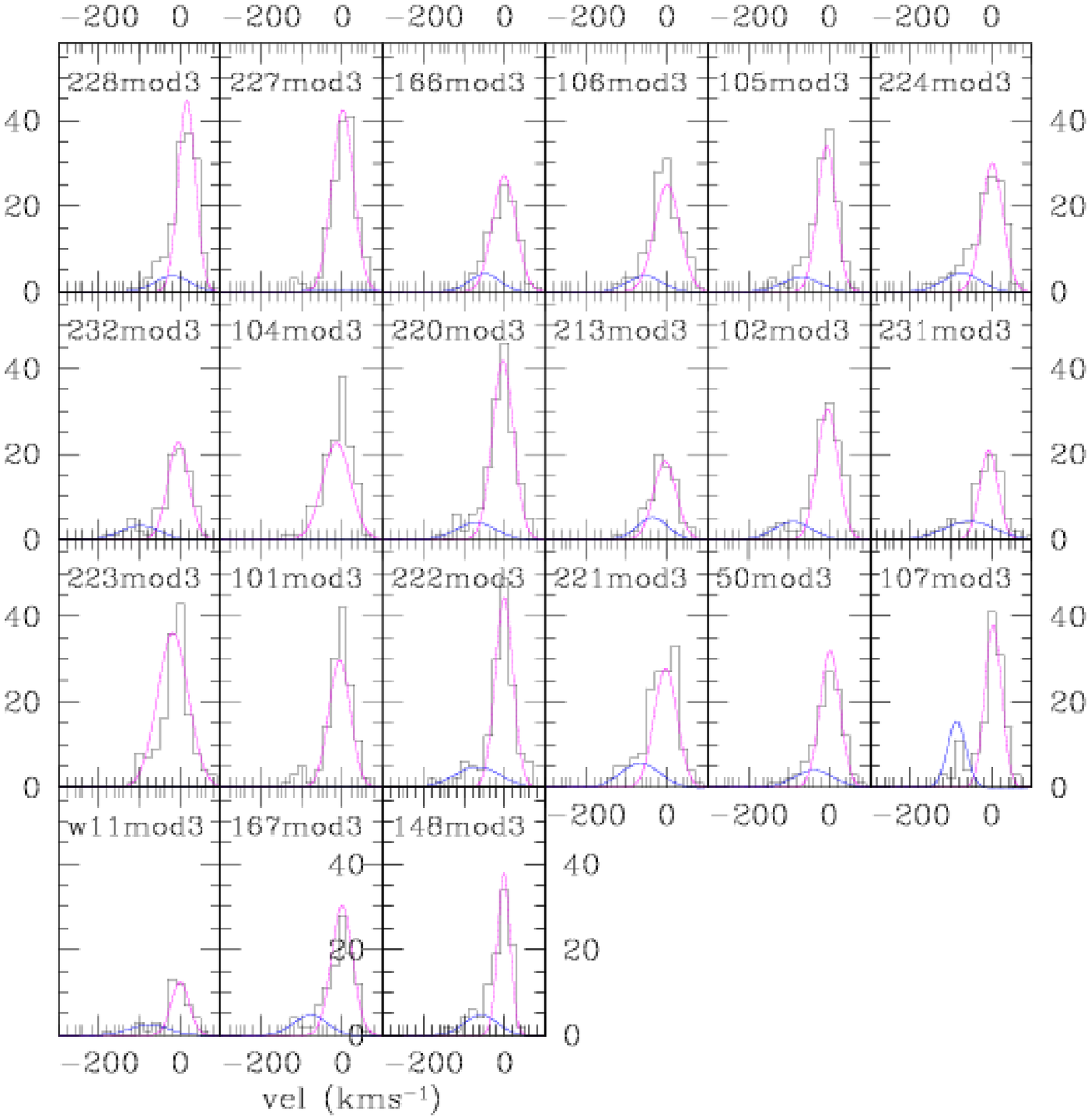}
\caption{A comparison of the data (top left panel) with 3 realisations
  of parsing our thin + thick disc model through the same analysis as
  our data, selecting stars from the same regions as the data. It can
  be seen that the model data resembles the actual data very closely,
  and that non-detections are likely an effect of sampling.  }
\label{models}
\end{center}
\end{figure*}

From this model data set, we select stars as we would select targets
to observe when designing DEIMOS masks, requiring them to have
$I$-band magnitudes between $22.0\ge I\ge20.5$. We then randomly
select the same number of stars as are observed at each field
location, and make velocity histograms in the disc-lag frame for each
field. We then analyse these distributions with the same GMM technique
described in \S~3.1, using a LRT to determine whether the distribution
of each model field is best fit by a single thin disc component, or a
double thin+thick component. This procedure is repeated 100 times,
allowing us to compute the average velocities and dispersions for each
component, plus sampling errors which we tabulate in
Table~\ref{mprops}. In our final 100 samples, the thick disc is
detected in 15 of the 21 fields on average. The fact that we do not
see the thick disc component in all our model fields implies that the
non-detections in our data are an effect of our sampling of the DEIMOS
fields rather than the component being absent in these fields. We show
the histograms and best fit Gaussians for three of these realizations
compared to our data in Fig.~\ref{models}.

\begin{table*}
\begin{center}
\caption{Average kinematic properties of model fields from 100 MC realisations}
\label{mprops}
\begin{tabular}{lccccc}
\hline
Field & v$_{thin}$ (disc frame, $\kms$) & $\sigma_{thin} (\kms)$ & v$_{thick}$ (disc frame, $\kms$) & $\sigma_{thick} (\kms)$ \\
\hline
228Mod & 11.4$\pm5.0$ & 19.2$\pm8.0$ & -50.0$\pm8.9$  & 38.3$\pm8.1$   \\
227Mod & -1.7$\pm0.6$ & 25.5$\pm10.8$& -37.1$\pm15.6$ & 46.3$\pm12.4$  \\
166Mod & -3.4$\pm2.1$ & 25.9$\pm7.3$ & -46.3$\pm9.1$  & 43.0$\pm13.4$  \\
106Mod & -4.4$\pm3.2$ & 25.9$\pm8.2$ & -54.3$\pm14.1$ & 42.8$\pm9.3$   \\
105Mod & -2.1$\pm1.2$ & 22.0$\pm9.4$ & -54.6$\pm13.8$ & 39.4$\pm12.8$  \\
224Mod & -2.7$\pm1.9$ & 23.1$\pm9.9$ & -40.2$\pm9.2$  & 42.3$\pm10.3$  \\
232Mod & -3.1$\pm2.7$ & 21.8$\pm8.2$ & -45.9$\pm12.2$ & 41.2$\pm13.2$  \\
104Mod & -2.5$\pm1.7$ & 22.8$\pm10.2$& -57.6$\pm14.4$ & 43.0$\pm11.0$  \\
220Mod & -7.0$\pm3.8$ & 21.1$\pm10.3$& -62.9$\pm8.7$  & 36.2$\pm10.4$  \\
213Mod & -8.1$\pm4.3$ & 19.5$\pm9.9$ & -55.1$\pm13.2$ & 32.2$\pm13.9$  \\
102Mod & -3.5$\pm2.7$ & 20.6$\pm8.9$ & -49.1$\pm7.9$  & 34.9$\pm12.4$  \\
231Mod & -3.7$\pm2.2$ & 19.4$\pm7.4$ & -49.6$\pm12.3$ & 33.2$\pm9.3$   \\
223Mod & -1.8$\pm1.1$ & 20.8$\pm7.2$ & -56.1$\pm11.7$ & 36.7$\pm8.8$   \\
101Mod &  1.9$\pm1.3$ & 22.3$\pm7.0$ & -64.1$\pm11.2$ & 38.4$\pm10.8$  \\
222Mod & -0.1$\pm1.2$ & 19.9$\pm6.6$ & -51.3$\pm12.6$ & 34.8$\pm11.0$  \\
221Mod &  3.9$\pm2.3$ & 19.9$\pm8.6$ & -46.2$\pm7.0$  & 38.1$\pm11.1$  \\
50Mod  &  4.2$\pm2.5$ & 17.9$\pm8.2$ & -42.2$\pm15.5$ & 33.5$\pm12.6$  \\
107Mod &  5.1$\pm2.7$ & 17.5$\pm7.7$ & -57.1$\pm11.6$ & 40.4$\pm12.9$  \\
w11Mod &  4.7$\pm1.1$ & 19.5$\pm6.8$ & -63.1$\pm10.4$ & 43.4$\pm10.5$  \\
167Mod &  2.0$\pm0.7$ & 22.5$\pm5.7$ & -52.2$\pm8.6$ &  41.6$\pm9.7$  \\
148Mod & -1.2$\pm1.5$ & 25.2$\pm6.8$ & -53.2$\pm9.7$  & 36.5$\pm10.3$  \\
\hline
\end{tabular}
\end{center}
\end{table*}

\begin{figure}
\begin{center}
\includegraphics[angle=0,width=0.9\hsize]{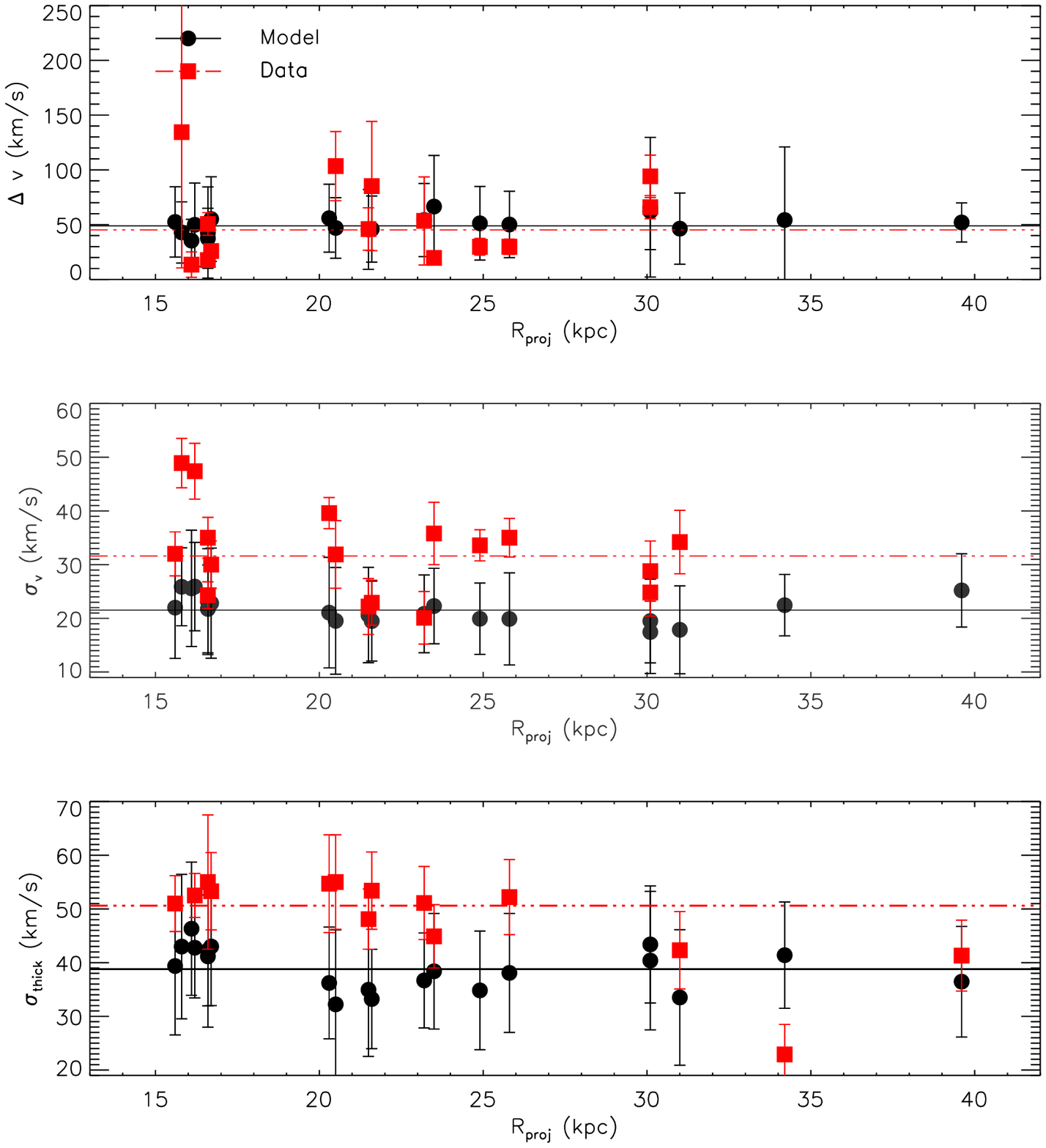}
\caption{This figure compares the results from our data with results
  from our model analysed in the same way. In all cases, data is
  represented by filled red squares and dot-dashed lines, and the
  model results are shown as filled black circles and solid
  lines. {\bf Top panel:} The difference in velocity, $\Delta v$,
  between the thin disc and thick component of data and model as a
  function of projected radius. The model lag is consistent with no
  evolution with radius, and shows an average lag of 48.9~$\kms$, very
  close to our input lag of 50~$\kms$. {\bf Middle panel: }Dispersion,
  $\sigma_{thin}$, of the thin disc is plotted for both data and model
  as a function of projected radius. The model thin disc is best fit
  with an average dispersion of 21.5~$\kms$, very close to the input of
  25.0$\kms$. {\bf Lower panel} Results for both data and model for
  the dispersion of the thick disc ($\sigma_{thick}$) as a function of
  radius. For our model, the thick disc is consistent with no
  evolution with radius, unlike our data, and has an average
  dispersion of $\sigma_v=38.8\kms$, which recovers our input
  dispersion of 40~$\kms$ relatively well. }
\label{modsummary}
\end{center}
\end{figure}

We now assess how both the lag between components and the dispersions
of each component evolve with radius for our model data set, and how
accurately we can recover these values from our model. In
Fig.~\ref{modsummary}, we plot these values for our model (black
circles) alongside the values we obtained from our data (red squares),
and fit the evolution of the model results with linear functions as
before. In the top panel, we show the measurement of $\Delta v$ for
our model fields as a function of projected radius. The model results
show no evolution of $\Delta v$ with radius, recovering an average lag across
all fields of 48.9$\pm6.7\kms$, which is similar to the constant lag
of 50~$\kms$ implemented in our model.

\begin{table}
\begin{center}
\caption{Average densities for thin and thick discs in model fields from 100 MC realisations}
\label{moddens}
\begin{tabular}{lccccc}
\hline
Field & $\rho_{* thin}$ (*/arcmin) & $\rho_{* thick}$ (*/arcmin) \\
\hline
228Mod   &73.0$\pm$35.2  &    31.0$\pm$    19.2 \\
227Mod   &15.7   $\pm$    7.1   &    18.4   $\pm$    8.1 \\
166Mod   &14.2   $\pm$    7.7   &    14.1   $\pm$    7.7 \\
106Mod   &11.9   $\pm$    5.7   &    11.8   $\pm$    5.7 \\
105Mod   &13.5   $\pm$    6.3   &    10.8   $\pm$    5.3 \\
224Mod   &10.6   $\pm$    5.0   &    10.1   $\pm$    4.8 \\
232Mod   &10.9   $\pm$    6.1   &    7.6    $\pm$    4.5 \\
104Mod   &10.7   $\pm$    4.9   &    7.9    $\pm$    3.8 \\
220Mod   &4.7    $\pm$    1.8   &    4.2    $\pm$    1.6 \\
213Mod   &4.6    $\pm$    2.6   &    4.4    $\pm$    2.4 \\
102Mod   &3.7    $\pm$    1.5   &    3.8    $\pm$    1.5 \\
231Mod   &3.8    $\pm$    1.9   &    2.8    $\pm$    1.3 \\
223Mod   &2.7    $\pm$    1.0   &    2.7    $\pm$    0.9 \\
101Mod   &2.9    $\pm$    1.1   &    2.8    $\pm$    1.0 \\
222Mod   &2.1    $\pm$    0.7   &    2.1    $\pm$    0.6 \\
221Mod   &1.7    $\pm$    0.5   &    1.9    $\pm$    0.5 \\
50Mod    &1.3    $\pm$    0.3   &    1.2    $\pm$    0.2 \\
107Mod   &1.4    $\pm$    0.3   &    1.5    $\pm$    0.3 \\
w11Mod   &1.4    $\pm$    0.4   &    1.5    $\pm$    0.3 \\
167Mod   &0.7    $\pm$    0.4   &    0.5    $\pm$    0.4 \\
148Mod   &0.6    $\pm$    0.4   &    0.5    $\pm$    0.3 \\
\hline
\end{tabular}
\end{center}
\end{table}

\begin{figure}
\begin{center}
\includegraphics[angle=0,width=0.99\hsize]{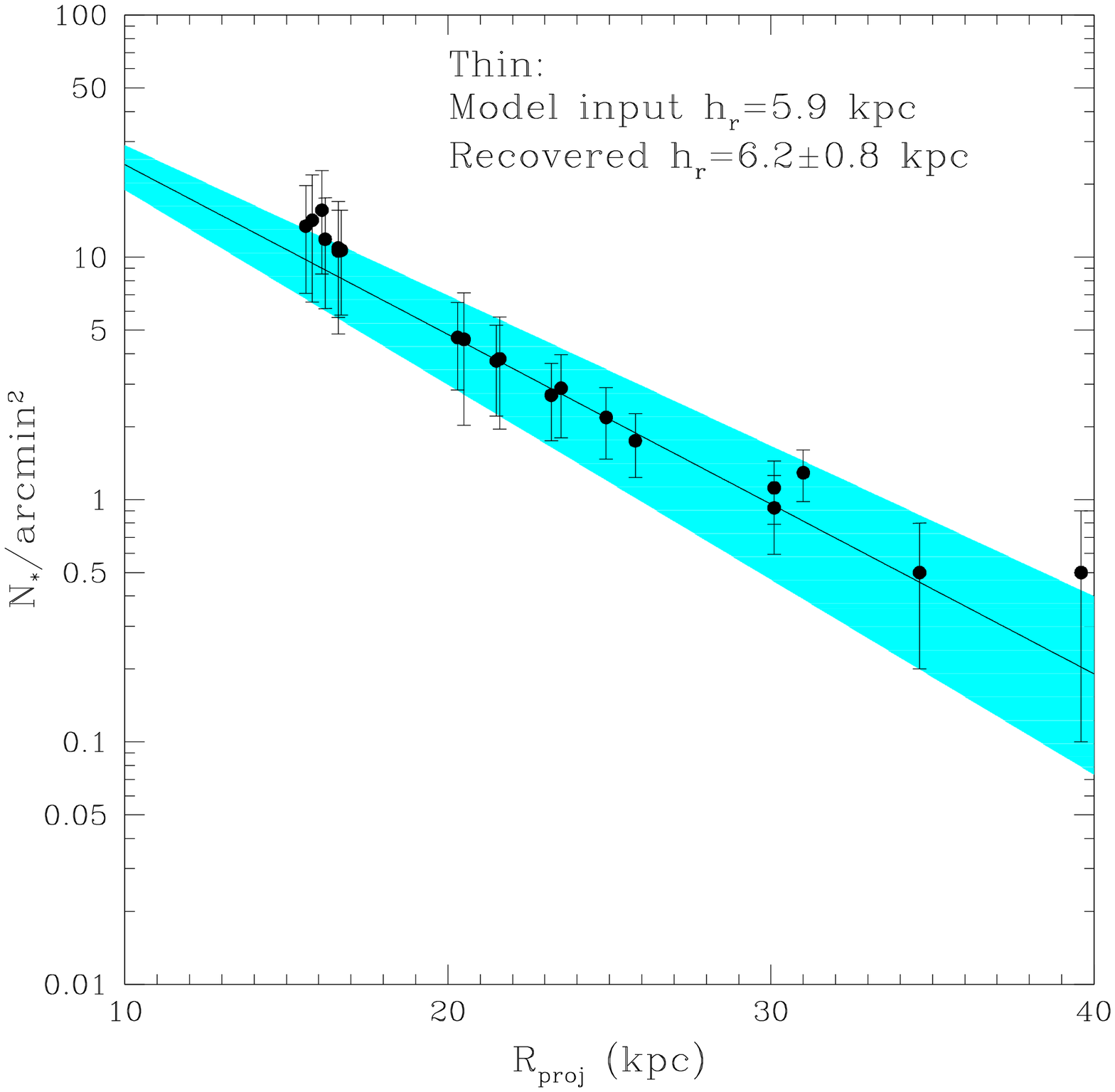}
\includegraphics[angle=0,width=0.99\hsize]{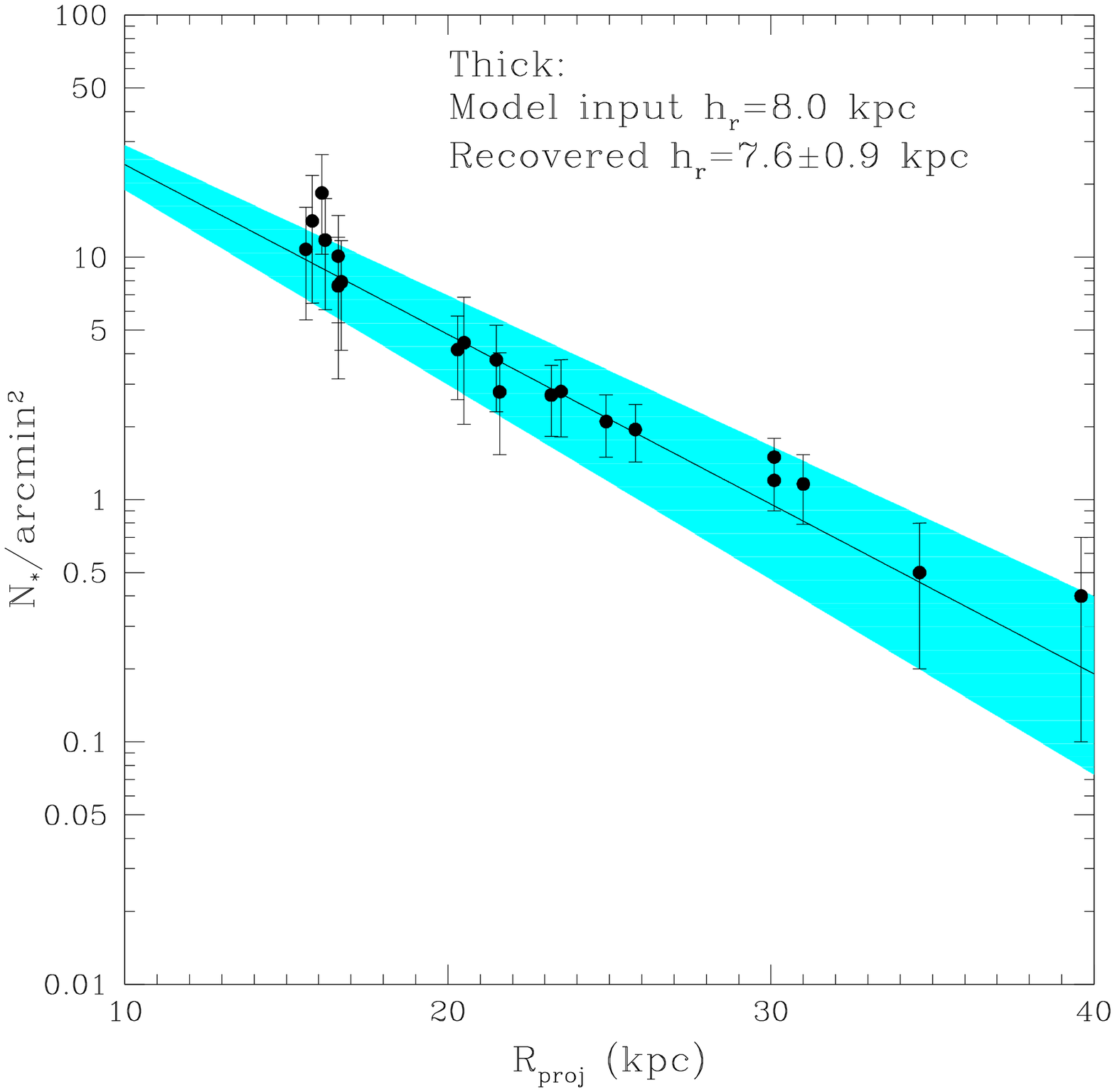}
\caption{Results from MC recovery of the scale lengths in our
  thin+thick disc model for our input thin (top panel) and thick
  (bottom panel) discs. The error bars on individual points represent
  the dispersion of calculated densities in the MC analysis, while the
  shaded regions represent the 1$\sigma$ uncertainties from the
  weighted least-squares fit. We recover a scale length for the thin
  disc of $h_r$=6.2$\pm0.8\kpc$ and h$_r=7.8\pm0.9$ for the thick
  disc, which are consistent with our input values.}
\label{modscale}
\end{center}
\end{figure}

Next, we compare the evolution of the disc dispersions for our model
with the data, shown in the central and lower panels of
Fig.~\ref{modsummary}. The model thin disc is best fit with a constant
relation, giving an average lag of $\sigma_{thin}=21.5\pm1.7\kms$,
very close to the input of 25.0$\kms$. For the thick disc dispersion,
$\sigma_{thick}$ the measured dispersion in the model
scatters about a mean dispersion of $\sigma_{thick}=38.8\pm2.4\kms$,
with no evidence of evolution with radius. 

Finally, we can use our model to get a handle on how accurate our
estimates of the scale length of the M31 discs might be. We use the
same Monte Carlo (MC) technique above to calculate the density of stars in each
component in our model fields 100 times, then we compute the average
density from these results. These results are shown in
Table~\ref{moddens}, and the errors represent the dispersion of the
densities computed in each field. We then plot the densities as a
function of radius for the thin and thick discs (shown in
Fig.~\ref{modscale}), and fit the result with an exponential profile
to determine the scale length. For the thin disc, we calculate
h$_r=6.2\pm0.8\kpc$, which is consistent with our input of 6.6 kpc,
and for the thick disc we compute h$_r=7.8\pm0.9\kpc$, consistent with
our input of 8.0 kpc. The shaded regions indicate the 1$\sigma$
uncertainties from the fit. These results suggest that our
observationally derived scale lengths for the thin and thick discs are
a good indicator of their true scale lengths.

\subsection{Comparison to the MW and `edge-on' thick discs}

Now that we have characterised the radial profile, kinematics and
metallicity of the thick disc in M31, we are able to compare it to the
properties of other thick discs that have been observed in the
universe. We shall begin with the most well studied thick disc
currently known -- that of our own Galaxy. Given that these two
galaxies are relatively close to one another (separated by 785 kpc),
and have similar morphologies (both large spiral galaxies),
comparisons between the MW and M31 are often made. But for all their
apparent similarities, these two galaxies are quite different from one
another. Work by \citet{hammer07} has shown that the MW is quite
different in terms of its structure and evolutionary history from the
majority of local spiral galaxies, whereas M31 is actually quite
``typical'', so these differences are perhaps unsurprising. In this
work, we have demonstrated that the scale lengths of the M31 discs are
larger than those of the MW by a factor of $\sim2$, as shown in
Fig.~\ref{height}. Given that we derive the scale heights of the M31
discs from these scale lengths, this results in scale heights in M31
that are of order $\sim3$ times as thick as those of the MW. However,
we note that as we calculate scale heights for the M31 disc based on a
relation determined from disc galaxies that are quite different in
terms of their mass to both the MW and M31, our values may be an
overestimate. The M31 discs are also seemingly hotter than the MW
discs, with $\sigma_{thin,M31}=32.0\kms$ cf.
$\sigma_{thin,MW}=20.0\kms$ \citep{ivezic08}, and
$\sigma_{thick,M31}=45.7\kms$ cf. $\sigma_{thick,MW}=40.0\kms$
\citep{ivezic08}. This could tell us something about the merger
history of M31. If the thick discs in both galaxies are formed as a
result of heating by mergers, the hotter discs of M31 could imply that
this galaxy has undergone a more active merger history than the MW.

The MW thick disc is more metal poor, enriched in $\alpha$ metals and
older than the thin disc. While we are unable to measure the age and
$\alpha$ abundances of the M31 discs, we have shown that there exists
an offset in the average metallicities of the two components of
$\sim0.2$ dex when measured spectroscopically. While we do not see
this offset photometrically, this could be due to our analysis
technique as we use isochrones of the same $\alpha$ abundance
([$\alpha$/Fe]=+0.2) and age (8 Gyrs) for both components. If we
modify the $\alpha$-abundance and age of these isochrones to
[$\alpha/$Fe]=+0.4 and 10 Gyrs for our thick disc sample, we see an
offset of $\sim0.2$ dex. We also note that both the thin and thick
discs in M31 appear to be more metal-poor than the MW discs, which
have average metallicities of [Fe/H]$\sim-0.3$ and [Fe/H]$\sim-0.6$
\citep{gilmore02,abadi03,carollo10} respectively, although there is a
significant metallicity spread in both discs. \citet{carollo10} also
demonstrated evidence of a secondary, more metal poor thick disc in
the MW, whose metallicities span the range $-1.8\le[Fe/H]\le-0.8$,
peaking at [Fe/H]=-1.3. This component also appears to be hotter than
the traditional MW thick disc component, with $\sigma_z$=44$\pm3\kms$,
very similar to what we observe in M31.

In \S4.1.3, we inferred scale heights for the thin and thick discs of
M31 of $h_z=1.1\pm0.2$~kpc and $h_z=2.8\pm0.6$~kpc respectively, using
a sample of 34 galaxies with thick discs measured by YD06 to determine
a relationship between scale length and scale height of a stellar
disc. As we noted in \S4.1.3, a comparison with the YD06 sample might
not be desirable, as these galaxies are typically much less massive
than M31, and selected to be bulgeless. A more appropriate comparison
would be the MW analogue, NGC 891, an edge-on galaxy that was recently
the subject of a structural analysis by \citet{ibata09} using HST/ACS
imaging. They detected the presence of a thick disc component in the
galaxy and were able to measure both a scale length and height for
this component of $h_r=4.8\pm0.1\kpc$ and $z_0=1.44\pm0.03\kpc$,
compared with $h_r=4.2\pm0.01\kpc$ and $z_0=0.57\pm0.01\kpc$ for the
thin disc component in this galaxy. This gives a ratio of $\sim1.1$
between the scale lengths and $\sim2.5$ for the scale heights of these
components, which is identical to what we observe in M31. To
illustrate this, we overplot these values for NGC 891 in
Fig.~\ref{scale} as a green circle.

In \citet{yoachim08a} the authors present kinematics of the thin and
thick discs of 9 of their initial sample of 34 galaxies, obtained
using the GMOS spectrograph on Gemini. To measure velocities and
dispersions in both thin and thick components, they placed slits in
positions corresponding to the midplane of the galaxy to measure the
thin disc properties, and above the midplane where the contribution
from the thin disc was thought to be negligible. As their typical
velocity resolution was 60$\kms$, they were unable to draw robust
conclusions on the velocity dispersions of these components, but they
were able to measure velocity rotation curves for each component, and
found a wide variety of behaviour amongst their thick disc components,
with discs which lagged behind the thin disc by only $\sim5\kms$,
discs that show no evidence of rotation and one case where the thick
disc is counter-rotating with respect to the thin disc. The average
lag between the thin and thick components of $\Delta v=46.0\kms$ we
see in the M31 system is larger than the majority that they
observe. We note that the galaxies in their sample were typically of
much lower mass than M31 (V$_{circ}<150\kms$ cf
V$_{circ}\sim230\kms$). In the most massive of their sample (which are
still less massive than M31), they do not detect a lag in the thick
disc kinematics at all, and they attribute this to contrast
issues. Their sample were also selected to be ``bulgeless'', unlike
M31 which has a significant bulge, and so a direct comparison may not
be advisable. Owing to the wide range of kinematic behaviour exhibited
in their sample, they conclude that the dominant formation process of
thick discs is via minor mergers and accretions of satellites. In
\citet{yoachim08b}, they use Lick indices to measure ages and
metallicities in 9 low mass galaxies with thick disc components. While
we measure an offset of 0.2 dex in the metallicities of the M31 thick
and thin discs, they were unable to measure any such offset in their
sample, though this could be a result of the insensitivity of Lick
indices to such differences at low metallicity. They do find that the
thick discs are host to older stellar populations than the thin disc,
however with our current data set, we are unable to comment on the
ages of stars in the M31 discs.

\subsection{Possible formation scenarios}

In this section, we discuss the various formation scenarios mentioned
in \S~1. Owing to our inability to measure ages and vertical dispersions in
M31, we are not able to confirm or reject any of these formation
mechanisms at present, so we discuss additional constraints for these
models that could help to rule out or confirm each scenario with
further data and analysis.

\subsubsection{Heating by minor mergers}

Numerous studies have identified that impacts and mergers of
satellites with masses less than a third of their hosts can
kinematically heat the thin stellar disc, puffing it out into a
substantially thicker disc
(e.g. \citealt{quinn93,robin96,walker96,velazquez99,chen01,sales09,villalobos09}). M31
is known to have recently undergone at least one significant minor
merger event, resulting in the GSS tidal stream. In recent work by
\citet{purcell10}, the authors model the heating of the stellar discs
by minor mergers and trace disc stars ejected into the stellar halo by
these simulated events. In addition to the stars ejected into the
halo, they observe a concomitant increase in the number of stars
located in the kinematic regime of the thick disc, contributing
$\sim~10-20$\% of the total stellar density along the major axis,
similar to what we observe in M31. They also find that their simulated
planar infall produces two-component systems with scale heights
(z$_{thin}\sim 1~\kpc$ and z$_{thick}\sim3~\kpc$), consistent with our
measurements for M31. The similarities between our findings and those
of \citet{purcell10} could suggest that thin disc stars heated by the
merging event that created the GSS may contribute some non-trivial
fraction of stars to the thick disc. 

 According to the simulations of \citet{kazantzidis09}, thick discs
 produced in this vein imprint a number of dynamical signatures on
 both the kinematic and structural properties of the galaxy. These
 include considerable thickening and heating at all radii, prominent
 flaring, particularly in the outskirts of the disc (beyond 3 scale
 lengths), surface density excesses at large radii, radial
 anisotropies and substantial tilting of the disc. As M31 is not edge
 on, we are unable to comment on the evolution of the height of the
 thick disc with radius, and so we cannot use this as a measure of
 flaring in the outer regions of the disc. However, one might expect
 that if there was a substantial flaring beyond 3 disc scale lengths
 ($\sim24$~kpc), that this may be reflected by an increase in the
 velocity dispersions of both thin and thick disc components. Our
 results for evolution in the thin and thick disc dispersions remain
 inconclusive, and so it is possible that such flaring may exist. At
 present, we possess few fields between R$\sim32$ and 39.6~kpc, so
 populating this region with kinematics, as well as additional fields
 further out, may further enlighten us to any potential
 flaring. Another test of this formation scenario would be to include
 fields from both the minor axis and NE portion of M31 to test for any
 radial anisotropy, assuming one can reliably disentangle
 contamination from foreground and substructure from the signatures of
 the discs. The work of \citet{sales09} also tells us that thick discs
 that are produced as a result of heating present structures with low
 orbital eccentricity.

\subsubsection{Accretion of satellite on a coplanar orbit}

Numerical simulations by \citet{abadi03} and \citet{penarrubia06} show
that an old, thick disc of stars could form via the accretion of stars
from satellite galaxies on an approximately coplanar orbit with its
host. Such discs are similar in radial extent and contain older
stellar populations when compared to the thin disc. The thick disc we
find in M31 is consistent with this model in so far as the radial
extents of both discs (5.9--7.3~kpc and 8.0~kpc) are comparable with
one another. They also argue that the mass and luminosity of the
progenitor satellite can be inferred from the metallicity of the
component. We deduce [Fe/H]$_{thick}=-1.0\pm0.1$ for M31, which would
correspond to a satellite of $M_V\sim-15$ ($L_v\sim9\times10^7\lsun$),
similar to the M31 dwarf elliptical, NGC 147 ($M_V=-15.1$,
\citealt{vandenbergh99}). However, given the mass we calculate for the
thick disc in \S4.1.4 of 2--4$\times10^{10}\msun$, it seems very
unlikely that the thick disc of M31 could have been formed from such a
satellite.

Results of the simulations of \citet{sales09} show that stars accreted
into a thick disc from satellites on coplanar orbits exhibit high
eccentricity orbits. Our present data set does not allow us to probe
the eccentricity of the orbits within the thick disc at this
time. With a larger data set, we could perhaps see the effects of
orbital differences in the form of structural asymmetries.

\subsubsection{Radial migration and internal heating}

The scattering of stars by spiral structure and molecular clouds has
long been proposed as a method of heating the stellar disc, moving
stars out onto more eccentric and inclined orbits
\citep{sellwood02,haywood08,roskar08,schonrich09a}, and it has been
argued in \citet{schonrich09a,schonrich09b} that these processes
naturally produce an old, $\alpha$-enhanced thick disc, whose
properties are consistent with those observed in the MW. These models
also show wide MDFs and an increase in the scatter of the
Age-Metallicity relation. This is also demonstrated in
\citet{quillen09}, where they investigate radial mixing induced by an
orbiting subhalo. Again they find evidence of wide MDFs in both the
thin and thick discs. With deeper photometry that allowed us to reach
the MSTOs of the two discs we could derive the average ages of these
components, and high resolution spectroscopy (R$\sim15,000$) of M31
thick disc stars that would allow us to determine accurate abundances
from unblended Fe lines for individual stars, we could comment more
robustly on the likelihood of such a formation scenario.

\subsubsection{Thick disc forms thick}

\citet{kroupa02} posited that thick discs could be formed as a result of 
vigorous star formation in massive star clusters ($\sim10^5-10^6\msun$) 
during the period of assembly of the stellar disc. If this is true, 
a number of these massive clusters may have survived to the present day, 
and would possess large vertical velocity dispersions. \citet{kroupa02} 
suggests that these clusters could be the metal-rich globular cluster system 
in the MW. Once again, owing to the inclination of M31, we are unable to 
measure the vertical dispersions of its metal-rich globular cluster system, 
and can therefore neither confirm nor reject this formation model.

\section{Conclusions}

Using the DEIMOS spectrograph on the Keck II telescope, we have
identified a statistically significant population of stars in M31 that
lags behind the thin and extended discs by 46.0$\pm3.9\kms$. Comparing
this with a model of a thin+thick disc system with the same distance
and inclination as M31 shows this component to be consistent with a
thick disc component. Analysing its kinematics, we find it to be
hotter than the thin disc, with average dispersion
$\sigma_{thick}=50.8\pm1.9\kms$ cf. $\sigma_{thin}=35.7\pm1.0\kms$,
larger than the dispersions observed in the MW discs. From composite
spectra for each component, constructed from highly probable thin and
thick disc stars (selected using stringent Gaussian cuts) we measure a
metallicity offset of $\sim0.3$ dex between the two disc, with the
thick disc being metal-poor than the thin disc
($\feh_{thick}=-1.0\pm0.1$ cf. $\feh_{thin}=-0.7\pm0.05$). The fact
that this metallicity offset is not observed when analysing the thin
and thick disc RGB stars with isochrones of identical age and
$\alpha$-abundance suggests that the two populations differ in these
properties, with the thick disc likely being older and more enriched
in $\alpha$ elements.

We measure scale lengths for both thin and thick discs, finding
$h_r=8.0\pm1.2\kpc$ for the thick disc, and $h_r=7.3\pm1.1\kpc$ for
the thin disc, comparable to previous estimates. Using the data of
YD06 we infer scale heights for both discs at $z_0=2.8\pm0.6\kpc$ and
$z_0=1.1\pm0.2\kpc$ for thick and thin discs respectively. These
values are of order 2--3 times larger than those measured in the MW,
perhaps suggesting that M31 has undergone more heating than our
Galaxy. 

By measuring the ratio of the densities of both discs, we are able to
estimate a mass range for the thick disc component of
2.4$\times10^{10}\msun<M_{*,thick}<4.1\times10^{10}\msun$. This value
provides a useful constraint on possible formation mechanisms, as any
proposed method for forming a thick disc must be able to heat (or
deposit) at least this amount of material.

Owing to current limitations within our data set, we are not able to
distinguish between the different thick disc formation
mechanisms. However, with further analysis of this component using our
complete kinematic sample (including regions from the minor axis and
NE of M31) and spectroscopic follow up of fields where this component
is strongly observed, we will be able to better understand the
chemistry of this component and distinguish between various formation
mechanisms.

\section{Acknowledgements}

MLMC would like to acknowledge the support of an STFC studentship. GFL
acknowledges the award of the Cambridge Raymond \& Beverly Sackler
distinguished visitor fellowship. 

\bibliography{mnemonic,michelle}{} \bibliographystyle{mn2e}

\end{document}